\documentclass[draft]{agujournal2019}
\usepackage{url} 
\usepackage{lineno}
\usepackage[inline]{trackchanges} 
\usepackage{soul}

\usepackage{amssymb}
\usepackage{amsmath}
\usepackage{graphicx}

\DeclareMathOperator{\atantwo}{atan2}

\nolinenumbers
\draftfalse

\journalname{JGR: Atmospheres}

\begin{document}

%
%

\title{Ozone Anomalies in Dry Intrusions Associated with Atmospheric Rivers}

%
%




\authors{Kirsten R. Hall\affil{1}, Huiqun Wang\affil{2}, Amir H. Souri\affil{3,4}, Xiong Liu\affil{2}, Kelly Chance\affil{2}}

\affiliation{1}{Center for Astrophysics $\vert$ Harvard \& Smithsonian, Radio \& Geoastronomy Division}
\affiliation{2}{Center for Astrophysics $\vert$ Harvard \& Smithsonian, Atomic and Molecular Physics Division}
\affiliation{3}{NASA Goddard Space Flight Center, Atmospheric Chemistry and Dynamics Laboratory}
\affiliation{4}{GESTAR II, Morgan State University}

\affiliation{1,2}{60 Garden St., Cambridge, MA 02138, USA}
\affiliation{3}{Greenbelt, MD, USA}
\affiliation{4}{Baltimore, MD, USA}




\correspondingauthor{Kirsten Hall}{kirsten.hall@cfa.harvard.edu}




\begin{keypoints}
\item Case studies and December climatology using MERRA-2 reveal positive tropospheric ozone anomalies within dry intrusions associated with ARs.
\item Average excess ozone concentrations are 10-13~ppbv at 400~hPa, and are even greater for increasing intensity of ARs.
\item STT of ozone associated with ARs in the NE Pacific may account for (13~$\pm$~2)\% of the total December Northern Hemisphere STT ozone flux.
\end{keypoints}

%
%

%
%


\begin{abstract}
As a result of their important role in weather and the global hydrological cycle, understanding atmospheric rivers’ (ARs) connection to synoptic-scale climate patterns and atmospheric dynamics has become increasingly important.
In addition to case studies of two extreme AR events, we produce a December climatology of the three-dimensional structure of water vapor and $\mathrm{O}_{3}$ (ozone) distributions associated with ARs in the northeastern Pacific from 2004-2014 using MERRA-2 reanalysis products.
Results show that positive $\mathrm{O}_{3}$ anomalies reside in dry intrusions of stratospheric air due to stratosphere-to-troposphere transport (STT) behind the intense water vapor transport of the AR.
In composites, we find increased excesses of $\mathrm{O}_{3}$ concentration, as well as in the total $\mathrm{O}_{3}$ flux within the dry intrusions,  with increased AR strength.
We find that STT $\mathrm{O}_{3}$ flux associated with ARs over the NE Pacific accounts for up to 13\% of total Northern Hemisphere STT $\mathrm{O}_{3}$ flux in December, and extrapolation indicates that AR-associated dry intrusions may account for as much as 32\% of total NH STT $\mathrm{O}_{3}$ flux.
This study quantifies STT of $\mathrm{O}_{3}$ in connection with ARs for the first time and improves estimates of tropospheric ozone concentration due to STT in the identification of this correlation.
In light of predictions that ARs will become more intense and/or frequent with climate change, quantifying AR-related STT $\mathrm{O}_{3}$ flux is especially valuable for future radiative forcing calculations.

\end{abstract}


\section*{Plain Language Summary}
Long filaments of rapidly moving water vapor in the atmosphere, known as atmospheric rivers (ARs), play a vital role in the Earth's water cycle. Because of this, research continues to expand into ARs' relationship with large-scale climate patterns. In this paper, we use data from the Modern Era Retrospective analysis for Research Applications to examine several extreme ARs that made landfall on the U.S. West Coast and their relationship to the transport of ozone from the stratosphere to the troposphere. We then combine eleven years of December AR and ozone data in order to study the average trend of ozone transport in connection with ARs. We quantify the AR-related ozone transport for the first time, and we find ARs with more intense water vapor transport result in the transport of higher concentrations of ozone. Quantifying ozone transport into the troposphere in connection with ARs is important as ARs may become more intense and/or more frequent with climate change, and ozone in the troposphere has consequences for the greenhouse effect.

%
%

%


%
%
%
%

\section{Introduction}

Atmospheric rivers (ARs) are streams of water vapor in the lower troposphere that are typically defined as being at least 2000~km in length and $\lesssim$1000~km in width.
They are often associated with warm conveyor belts of ascending moist air travelling along and ahead of the cold fronts of extratropical cyclones (ECs) \cite{dacr15,Eira18,Guo20}; though about 20\% occur not in connection with an EC, and neither the intensity nor the precise location or duration of the AR can be determined from the cyclone \cite{zhan19}.
They are the primary mode by which water vapor is transported from the tropics towards the poles \cite{Newe92,Newe94,Zhu98}.
They occur all over the globe, and there is a high concentration of recent studies focused on expanding the global characterization and analysis of ARs \cite<e.g.,>{Shie18, Coll20, Coll22}.

The intensity of ARs, as defined by the integrated vapor transport (IVT), is variable, and strong AR events can have catastrophic consequences, such as flooding and mudslides.
Throughout this analysis, we investigate composites of IVT and ozone profiles connected to ARs with a range of intensities with IVT$\geq$250~kg~m$^{-1}$~s$^{-1}$ and a range of duration from less than 24 hours to more than 72 hours.
\citeA{Ralp19} devises a categorization scheme for ARs as Cat 1-5, and we will use this as a reference in discussing some events.
Climate model projections using Phase 5 of the Coupled Model Intercomparison Project (CMIP5) indicate that their intensity (IVT strength) will increase in the warming future by up to 25\%, and their frequency will increase by anywhere from 30\% to 300\% globally, depending on the definitions of the number of AR days in different detection algorithms \cite{Gao15, Espi18}.
As a result of their important role in weather and the global hydrological cycle, understanding ARs' connection to planetary- and synoptic-scale climate patterns and atmospheric dynamics has become increasingly important.

Of particular interest is the impact of upper-tropospheric dynamics on ARs and vice versa.
For example, several studies link ARs and extreme precipitation events directly to Rossby wave breaking (RWB) \cite<e.g.,>{Payn14, Ryoo15, Payn16, Hu17,Moor19}.
In an assessment of the connection between intense moisture transport and RWB, \citeA{deVr21} finds that the depth of the wave breaking and intensity of the integrated vapor transport are both directly correlated with the severity of extreme precipitation.
That is, through the analysis of the depth of folding of PV contours, they find that the deeper the breaking, the more intense the water vapor transport. 

With a clear connection between moisture transport and RWB, it follows to investigate a connection to the transport dynamics from the stratosphere to the troposphere.
Stratosphere-troposphere transport (STT) or exchange (STE) of air masses on a planetary scale is a consequence of Rossby waves propagating into the stratosphere and mesosphere.
This phenomenon leads to both the rising of air from the troposphere to the stratosphere in the tropics, and a descent at mid- and high- latitudes \cite{Holt95, Nath16}.
PV anomalies or streamers exist as a result of dry intrusions of ozone-rich stratospheric air that lag behind the transport of moist air from the equatorial lower troposphere poleward and into the upper troposphere \cite{Waug05}.
These PV intrusions as a result of STE increase the tropospheric ozone ($\mathrm{O}_{3}$) concentrations \cite{Nath16}.

In addition to the planetary-scale continuous downward flow of air masses, in the extratropics, STE also occurs as an episodic phenomenon in association with synoptic-scale processes that perturb the tropopause.
For example, STE in connection with ECs has been well-documented \cite<e.g.,>{Dani68, John81, Holt95, Wern02, Reut15}.
Dry intrusions, or dry airstreams \cite{Brow97}, of stratospheric ozone-rich air are a defining characteristic in the cold sector of synoptic-scale low pressure systems, and they are also described by relatively high potential vorticity (PV) in a folding or lowering of the tropopause.
\citeA{Jaeg17} produces the climatology of positive $\mathrm{O}_{3}$ anomalies due to STE in dry intrusions associated with ECs, finding these events contribute up to 42\% of the total STE $\mathrm{O}_{3}$ flux.
We aim to provide in this work a similar assessment of STE of $\mathrm{O_3}$ in PV intrusions linked to December AR events in the northeast (NE) Pacific.

Understanding the total stratospheric contribution to tropospheric $\mathrm{O}_{3}$ concentrations has been a significant area of research for several decades \cite{Monk00,Stoh03,lefo11,Sker14,abal20}.
Tropospheric $\mathrm{O}_{3}$ is a greenhouse gas and key contributor to climate change \cite{Laci90, Word11}, and it is a pollutant at the surface, harmful to humans and plants.
$\mathrm{O}_{3}$ impacts the oxidation capacity of the atmosphere affecting the chemical pathway of formation and loss of many species with crucial consequences for radiative forcing.
Due to STE, there remain large discrepancies amongst chemical transport models' predictions of tropospheric $\mathrm{O}_{3}$, which impedes a reasonable understanding of the total $\mathrm{O}_{3}$ concentration due to photochemistry \cite{Youn13, Youn18, Morg18, Grif21}.
Disentangling the contributions to tropospheric $\mathrm{O}_{3}$ by STE from the production of $\mathrm{O}_{3}$ due to precursor emissions is vital for understanding air pollution and future warming due to greenhouse gases.

\begin{figure*}
\centering
    \includegraphics[width=5.5in]{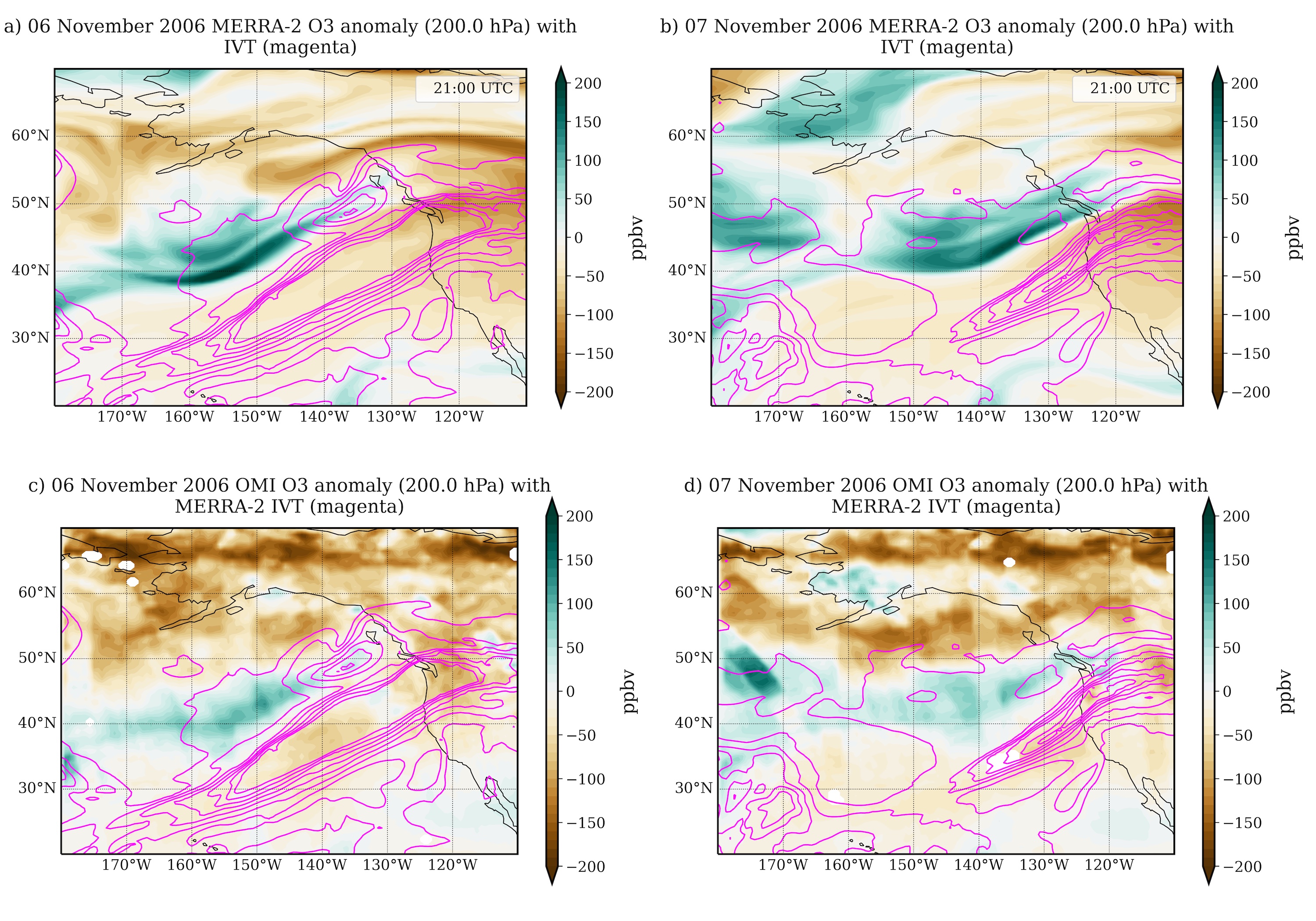}
\caption{MERRA-2 at 21:00 UTC (top: a,b) and OMI (bottom: c,d) anomalous $\mathrm{O}_{3}$ at 200~hPa (shading) for 06 (left) and 07 (right) November 2006. In all plots, the magenta contours are MERRA-2 IVT in steps of  100~kg~m$^{-1}$~s$^{-1}$, with the innermost contour 700~kg~m$^{-1}$~s$^{-1}$. }
\label{omi}
\end{figure*}

In the application study of water vapor retrieval from the Ozone Monitoring Instrument \cite<OMI,>{Leve06}, \citeA{Wang19} discovered anomalous $\mathrm{O}_{3}$ concentrations at the level of the tropopause trailing an exceptional (Cat 5) atmospheric river event on 06-07 November 2006 (Figure~\ref{omi}, bottom row).
This AR occurred over the northeast Pacific and brought detrimental flooding to the Pacific Northwest upon landfall.
The details of this event are described in \citeA{Neim08a}.
The $\mathrm{O}_{3}$ anomaly appears to the northwest of the AR, mirroring the curvature of the AR as a long filament of excess $\mathrm{O}_{3}$ relative to the November decadal climatology, both computed from the OMI $\mathrm{O}_{3}$ mixing ratio data interpolated to 200~hPa.
OMI data suggest paired anomalies in the upper tropospheric $\mathrm{O}_{3}$ and total column water vapor in this AR event.
This discovery inspired a detailed analysis of the time evolution of the three-dimensional structure of the $\mathrm{O}_{3}$ and water vapor distributions associated with representative AR events using the MERRA-2 reanalysis product.
While approximately 82\% of ARs in the North Pacific are formed in connection with ECs \cite{zhan19}, they are distinct phenomena of varying size, evolution, and duration relative to their cyclone counterparts, thus we aim to quantify the STE in relation to ARs and their physical properties.

In this paper, we uncover the relationship between ARs and associated anomalies in tropospheric $\mathrm{O}_{3}$ concentrations as a result of related stratosphere-troposphere exchange (STE).
The scope of this work focuses broadly on ARs over the NE Pacific and western North America; first, we give a detailed view of several ARs, then we present composites of December ARs over an eleven year period.
We describe the data in Section~\ref{data}, our analysis methods, including tracking ARs over their lifetime, and our composite analysis in Section~\ref{ana}.
Section~\ref{casestudies} details our assessment of anomalous $\mathrm{O}_{3}$ associated with two exceptional (Cat 5) \cite{Ralp19} AR events in November 2006 and December 2010, as well as a complete assessment of all ARs in December 2010.
We extend this analysis to all of December 2004-2014 with a statistical summary and composite results, including average IVT, $\mathrm{O}_{3}$ anomaly, and vertical cross sections in Section\ref{composite_results}. In this section, we also compute the average STT $\mathrm{O}_{3}$ flux and divide our sample into two bins of AR IVT to assess the correlation between excess $\mathrm{O}_{3}$ and AR intensity.
We summarize and conclude in Section~\ref{conclude}.

\section{Data}
\label{data}

\subsection{Ozone Monitoring Instrument (OMI)}

We validate our use of MERRA-2 $\mathrm{O}_{3}$ profiles by comparing MERRA-2 $\mathrm{O}_{3}$ mixing ratios and anomalies to those retrieved from OMI onboard the Aura spacecraft.
OMI is an imaging spectrometer that covers visible-UV wavelengths from 350-500~nm, and has been making daily global measurements since October 2004 at nominal nadir spatial resolutions of 13×48~km and 13x24~km for spectral windows UV1 and UV2, respectively, from a polar orbit with 13:30 Equator crossing local time.
The 10-year OMI Ozone Profile (PROFOZ) product used in this study was validated with ozonesondes \cite{Huan17} and Microwave Limb Sounder \cite{Huan18} measurements. Compared to ozonesondes observations, OMI profiles were found to have mean biases less than 6\% in the midlatitudes down to an average of 550~hPa \cite{Huan17}.
In comparison to MLS, the profiles are found to have negative global mean biases between 3-9\% below approximately 100~hPa \cite{Huan18}.
These biases are due to a combination of OMI retrievals and MLS having a positive bias at these altitudes compared to other measurements \cite{Froi08}.
The standard deviations are approximately 27\% and 20\% at these latitudes and pressure levels in comparison with ozonesondes and MLS, respectively \cite{Huan17, Huan18}.

Specifically, we use daily $\mathrm{O}_{3}$ profiles, first generated with a regridding of  0.5$^\circ$~x~0.5$^\circ$ resolution to improve the SNR and speed up production from \citeA{Liu10} for 06 and 07 November 2006.
We refer the reader to \citeA{Liu10}, \citeA{Huan17}, and \citeA{Huan18} for the full details of the profiles retrievals and validations.
\citeA{Wang19} used these data regridded at 1$^\circ$~x~1$^\circ$ in an application demonstrating the retrieval of Total Column Water Vapor (TCWV) from OMI for the AR event on these dates and discovered the high $\mathrm{O}_{3}$ mixing ratio concentrations parallel to the northwest of the excess TCWV from the AR.
We then regrid our 0.5$^\circ$ latitude by 0.5$^\circ$ longitude OMI data to the MERRA-2 resolution of 0.5$^\circ$ latitude by 0.625$^\circ$ longitude using the publicly available python bilinear method regridding algorithm xESMF.Regridder \cite{Zhua23} for our comparison in Section~\ref{2006}.

\subsection{Modern-Era Retrospective analysis for Research and Applications, Version 2 (MERRA-2)}

The atmospheric data products used in this analysis are taken or calculated from the Modern-Era Retrospective analysis for Research and Applications, Version 2 \cite<MERRA-2,>{Gela17} reanalysis, which covers the period of January 1980 through to within a couple weeks of real-time, at a horizontal resolution of 0.5 × 0.625° latitude-by-longitude.
MERRA-2 assimilates meteorological and $\mathrm{O}_{3}$ observations using the updated Goddard  Earth  Observing  System  Model (GEOS) model and data assimilation system \cite{Molo15,Gela17}.
In particular, we select data sets with an instantaneous 3-hourly temporal resolution with 42 pressure levels from 1000~hPa to 0.01~hPa, as well as the monthly mean instantaneous products, and use meteorological and chemical products for winds (u,v), temperature, specific humidity, Ertel's PV, relative humidity, sea level pressure, and $\mathrm{O}_{3}$ mixing ratios \cite{Merra2_3hr, Merra2_M}.
The GEOS model system for MERRA-2 uses the Goddard 2D chemistry and transport model to derive linearly interpolated $\mathrm{O}_{3}$ daily values for both the stratosphere and the troposphere \cite{Niel17,Know22}.

Prior to 2004, MERRA-2 assimilated $\mathrm{O}_{3}$ profiles from several NOAA Solar Backscatter Ultraviolet (SBUV/2) instruments. After 2004, and for years relevant to this work, MERRA-2 improves upon the ozone assimilation with total column ozone from OMI, as well as version 2.2 ozone retrieval from Aura's Microwave Limb Sounder \cite<MLS,>{Wate06} for which the recommended vertical range is 215~hPa to 0.02~hPa as described in \cite{Warg17}.
Ozone mixing ratios from MERRA-2 in the lower stratosphere agree with the Stratospheric Aerosol and Gas Experiment II measurements to within 5$\%\ $ in the extratropics, and once OMI and MLS data were assimilated, the bias drops to 2$\% $ \cite{Warg17}.
\citeA{Warg17} reports similar $\lesssim$5$\% $ bias in the extratropics ($\sim$30-60$^\circ$N) in comparison to ozonesondes observations of stratospheric ozone.
In the upper troposphere within the SBUV period, the bias is within 10$\% $ of the ozonesondes \cite{Warg15}.
In the Aura period after 2004, the low sensitivity of OMI columns to $\mathrm{O}_{3}$ mixing ratios for $P\lesssim$500~hPa along with the simple chemistry in the assimilation system \cite{Warg17} cause the MERRA-2 tropospheric $\mathrm{O}_{3}$ to be biased low by 13.6$\% $ on average in comparison to ozonesondes between 30-60$^\circ$N.
Given the agreement in the lower stratosphere, tropospheric measurements of $\mathrm{O}_{3}$ mixing ratio inside of dry, stratospheric intrusion events are expected to be reliable except for the consideration that the background $\mathrm{O}_{3}$ in the troposphere is biased due to the simple chemistry transport, which is inconsequential for this study.
\citeA{Know17} shows that the vertical structure of $\mathrm{O}_{3}$ in stratospheric intrusions over the western United States is well captured by MERRA-2, and that MERRA-2 has high enough horizontal resolution to capture $\mathrm{O}_{3}$ filaments \cite<as also demonstrated in>{Ott16}.
This analysis primarily identifies excess anomalous $\mathrm{O}_{3}$ in dry intrusions in the upper troposphere. Any assessment below roughly 500~hPa is trustworthy as it is within the stratospheric intrusion as similarly found in \citeA{Know17} and \citeA{Jaeg17}.

\subsection{AR Identification Catalog}

To identify ARs, we use the binary tag files from the Atmospheric River Tracking Method Intercomparison Project \cite<ARTMIP,>{Shie18} MERRA-2 Tier 1 catalogues \cite{Rutz19}.
ARTMIP is an international collaboration to quantify uncertainties in AR science based on the details of detection algorithms.
The Tier 1 catalogues are a set of AR catalogues all created using MERRA-2 data, but with differing detection criteria and implemented on varying geographical regions, global and otherwise.
Our primary criteria in this study for an AR identification catalog is that they cover the NE Pacific Ocean, classify ARs with a broad range of intensities -- as indicated by the integrated water vapor transport (IVT), and specifically IVT $\geq$ 250~kg~m~s$^{-1}$ -- and track them for the duration of their lifetime.
Moreover, using a catalog that identifies ARs globally is useful for future analyses of $\mathrm{O}_{3}$ in dry intrusions associated with ARs globally.
There are many methods for identifying ARs, and based on our criteria, we opted to primarily use the ARTMIP binary tag catalogue generated using the algorithm from \citeA{Rutz14}.

The \citeA{Rutz14} detection algorithm uses selection criteria that enforce the AR length to be at least 2000~km and the minimum absolute integrated water vapor transport (IVT) from surface to 100~hPa to be greater than or equal to 250~kg~m$^{-1}$~s$^{-1}$.
The catalog is global, and \citeA{Rutz14} provides an analysis of the climatology of the characteristics of ARs over the Western United States.
The algorithm used to detect ARs in \citeA{Rutz14} was originally run on the Interim European Centre for Medium-Range Weather Forecasts (ECMWF) Re-Analysis (ERA-Interim; \citeA{Dee11}) data.
We acquired the catalog from the NCAR Climate Data Gateway as a part of Tier 1 of the ARTMIP project, in which the \citeA{Rutz14} algorithm was run on the MERRA-2 3-hourly data to generate binary tags -- 0s for no AR in that grid cell and time step and 1s for AR present at that location and time.

In this work, we examine any relationship between the amplitude of $\mathrm{O}_{3}$ anomaly and $\mathrm{O}_{3}$ flux and AR strength. The \citeA{Rutz14} algorithm is conservative in its categorization of ARs, allowing us to sample the full range of AR IVTs (down to peak IVT of 250~kg~m$^{-1}$~s$^{-1}$).
As a check on the dependence of our results on our catalog, we ran our algorithm using several other ARTMIP Tier 1 participating developer AR identification catalogs for December 2010 and obtained the same results for each one.
The additional catalogs we tested used algorithms from: \citeA{Kash21} ClimateNet deep learning algorithm, and \citeA{Reid20} Reid250 with absolute IVT$>$250~kg~m$^{-1}$~s$^{-1}$.
We intentionally tested other algorithms for which the IVT values are allowed to be at least as low as 250~kg~m$^{-1}$~s$^{-1}$. The ClimateNet deep learning algorithm imposes no minimum IVT threshold, and the Reid250 algorithm imposes an additional length-width requirement.
There was no dependence of our results on the choice of AR binary tag catalog, likely because we only track one AR at a time as identified by the maximum IVT in the region, and so long as the selection criteria includes ARs with IVT$>$250~kg~m$^{-1}$~s$^{-1}$, then we are likely to identify at least the same ARs over our geographical area.
Other criteria could have imposed differing lifespans of ARs, but in spite of any differences we find out results unchanged.
If instead we were to choose an algorithm with a more strict IVT threshold, we would miss AR time steps that drop below that, but that would go against our intended purpose of comparing a wide range of possible AR intensities, so we did not test this.

\section{Analysis Methods}
\label{ana}

Here we detail the various components of our analysis.
The general algorithm progresses as follows: use MERRA-2 data to compute IVT in the NE Pacific, apply a mask of binary tags that identify ARs at each time step, track individual ARs through their lifetime, and simultaneously identify nearby $\mathrm{O}_{3}$ anomalies that are a result of dry intrusions.
We track the evolution of the ARs using the binary tags and peak IVT values at each time step, as well as the vertical structure of the associated dry intrusion using PV.
The peak IVT value, direction of the AR ($\Theta_{IVT}$), peak anomalous $\mathrm{O}_{3}$ concentration, and locations of the peaks are all stored for further analysis.
We also describe our calculation of $\mathrm{O}_{3}$ flux inside the dry intrusions and the computation of composite IVT and O3 anomalies.

\begin{figure}[!t]
 \centering
 \includegraphics[width=5.5in]{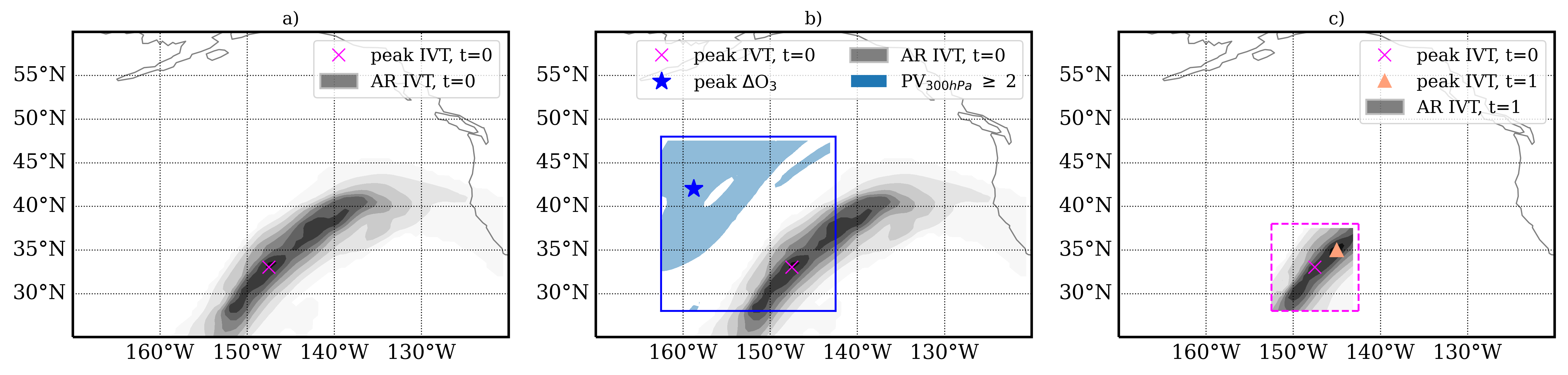}
 \caption{Visualization of AR tracking and identification of peak anomalous $\mathrm{O}_{3}$ on pressure surfaces using MERRA-2 data for 11 December 2010. Left panel, step one: identify the AR (gray shading) and its peak IVT (magenta x). Middle panel, step two: within 15$^\circ$~N, 15$^\circ$~W, 5$^\circ$~S, and 5$^\circ$~E of the location of the peak IVT (blue box), identify potential tropopause lowering as PV$\geq$2 at 300~hPa (blue shading), then find the peak value of anomalous O$_3$ (blue star) at 200 and 400~hPa. The location of the peak value at 400~hPa will be used along with the peak IVT location to explore the vertical cross section and further characterize possible tropopause folding. Right panel, step 3: within a 10 degree latitude by 10 degree longitude box (magenta dashed box) centered on the peak IVT at t=0 (magenta x), identify the peak IVT for the AR at the next time step, t=1 (peach triangle). The algorithm repeats steps two and three with peak IVT at the new time step replacing the previous value and location until the AR dissipates or a new, more intense AR enters the smaller search box.}
 \label{method}
\end{figure}

\subsection{Atmospheric river tracking}
\label{IVTana}

We built an algorithm to track ARs through their lifetime.
We make maps of vertically integrated water vapor transport (IVT, kg~m$^{-1}$~s$^{-1}$) computed as,
\begin{equation}
    \mathrm{IVT} = \frac{1}{\mathrm{g}} \sqrt{ ( \int  \mathrm{qu dp})^2 + (\int \mathrm{qv dp})^2}
\label{ivteq}
\end{equation}
where q is the specific humidity, and u and v are the zonal and meridional layer-averaged wind components, respectively, all of which are taken from MERRA-2 3-hourly datasets.
The integral runs over each layer's pressure increment, dp, and g is the acceleration due to gravity.
Using the binary tag catalog from \citeA{Rutz14} as a mask on the IVT maps to identify the locations of ARs, we find the peak amplitude of the IVT in the NE Pacific Ocean, from 180$^{\circ}$W to 110$^{\circ}$W and 25$^{\circ}$N to 60$^{\circ}$N, at each three hour time step beginning on 01 December at 00:00 for each year between 2004-2014.
We also run this algorithm on November 2006 data.
The decision to choose the maximum IVT means we are choosing the most intense AR in the field at the initial time step, but as the algorithm is designed to follow ARs until they dissipate, we still sample a wide range of AR intensities.
We also keep track of the AR direction of travel defined according to \citeA{Pan19} as,
\begin{equation}
    \Theta_{\mathrm{IVT}} = \atantwo (\mathrm{IVTu}, \mathrm{IVTv}),
\end{equation}
where $\mathrm{IVTu}$ and $\mathrm{IVTv}$ are the zonal and meridional IVT components defined by,
\begin{equation}
    \mathrm{IVTu} = \frac{1}{\mathrm{g}} \int  \mathrm{qu dp}
\end{equation}
and
\begin{equation}
    \mathrm{IVTv} = \frac{1}{\mathrm{g}} \int  \mathrm{qv dp}.
\end{equation}

Once the algorithm identifies a new AR, it follows the AR through its remaining lifespan by searching within a 10$^{\circ}$ latitude by 10$^{\circ}$ longitude box centered on the initial peak IVT location for the peak IVT at the next time step.
This is exemplified in steps one and three of Figure~\ref{method} (left and right panels).
When the peak IVT within this smaller field drops to zero, the search box returns to the original area encompassing the NE Pacific to search for the next AR event.
This enables us to determine at each 3-hourly time step the peak IVT of a single AR event throughout the majority of its lifetime.
The algorithm only tracks one AR at a time; therefore, it may miss the initial stages of an AR's development.
Additionally, if the peak IVT increases by more than 125~kg~m$^{-1}$~s$^{-1}$ compared to the previous time step, the algorithm checks for the peak IVT in the entire NE Pacific area because such an increase in IVT may indicate that a different AR has entered the search field.
If the location of that peak is more than 1000~km from the previous peak location or more than 450~kg~m$^{-1}$~s$^{-1}$ greater than the previous time step's IVT, then that new peak IVT is determined to be associated with a new AR.

\subsection{STT identification and characterization} 
\label{o3ana}

$\mathrm{O}_{3}$ enhancements in the UT/LS often result from the transport of dry, ozone-rich stratospheric air down toward and into the troposphere behind the cold front of the EC and/or AR.
The $\mathrm{O}_{3}$ enhancement can be associated with a complete folding or simply a lowering of the dynamic tropopause.
While some intrusions might be associated with both an EC and an AR if the AR is coincident with the EC, we are making a distinct measurement of intrusions associated with ARs.
ARs occur in coincidence with ECs approximately 82\% of the time \cite{zhan19}, but even in these instances, the $\mathrm{O}_{3}$ anomaly may be distinct from that associated with the EC and/or may evolve with the AR and persist for a longer duration of time \cite{Mund16, zhan19}.
For comparison, \citeA{Jaeg17} provides a complete analysis of the $\mathrm{O}_{3}$ in dry intrusions to the southwest of ECs in the Northern Hemisphere (NH).
Similarly, \citeA{Know15} quantify tropospheric $\mathrm{O}_{3}$ enhancement in the area encompassing the most intense (95$^{th}$ percentile) spring ECs over the North Pacific and North Atlantic.

In order to better understand the relationship between anomalous $\mathrm{O}_{3}$ and ARs, we track the excess $\mathrm{O_3}$ in proximity to several AR case studies and in composites.
We compute $\mathrm{O}_{3}$ anomalies by subtracting the relevant monthly climatology over the years 2004-2014.
We track the tropopause lowering or folding using the PV provided by MERRA-2.
We use the dynamical tropopause definition of whichever isosurface is lower: PV$\geq$2~PVU (1~PVU = 10$^{-6}$~K~m$^2$~kg$^{-1}$~s$^{-1}$), or 380K \cite{Hosk85,Holt95}.
After identifying the AR and its peak IVT, our algorithm identifies potential tropopause lowering using PV contours, PV$\geq$2 at 300~hPa within 15$^\circ$~N, 15$^\circ$~W, 5$^\circ$~S, and 5$^\circ$~E of the location of the peak IVT (Figure~\ref{method}, middle).
We then locate the peak anomalous $\mathrm{O}_{3}$ concentration within the dry intrusion at the 200~hPa and 400~hPa pressure levels.
Maximum $\mathrm{O}_{3}$ anomalies at these pressure levels are also required to lie outside the AR as determined by the binary mask as an additional means of identifying the tropopause fold or lowering.
Figure~\ref{method} demonstrates the process of identifying the AR, its peak IVT, and maximum anomalous $\mathrm{O}_{3}$ concentration at some time t=0, then shows the box that is search for the peak IVT at the next time step, t=1.
The algorithm repeats steps two and three with peak IVT at the new time step replacing the previous value and location until the AR dissipates or a new, more intense AR enters the smaller search box.

In order to track and characterize the depth of the dry intrusions and the STT of $\mathrm{O}_{3}$, we use the vertical pressure cross section over a geodesic line defined according to the peak IVT and the maximum $\mathrm{O}_{3}$ anomaly at 400~hPa.
Along this cross-section, we identify the dynamic tropopause, which in this case is always the PV=2 contour.
By tracking this contour, we follow the tropopause folding or lowering and record its maximum depth (highest pressure level reached).
We are careful to ensure the contour is connected to the stratosphere and is not part of a cutoff according to \citeA{Sker14}.
This is exemplified in the case study analyses in Section~\ref{casestudies} and the results are further explored in Section~\ref{composite_results}.

A further quantity of interest is the $\mathrm{O}_{3}$ flux due to STT.
Following a similar method to \citeA{Jaeg17}, we quantify the total $\mathrm{O}_{3}$ in the dry intrusion between the local 2~PVU dynamical tropopause and the level 75~hPa below the climatological dynamical tropopause, which we compute to be 275~hPa using the MERRA-2 PV data for the eastern part of the North Pacific in December 2004-2014 between 25$^{\circ}$~N to 60$^{\circ}$~N and 180$^{\circ}$~W to 110$^{\circ}$~W.
This computation puts the upper bound on the dry intrusion at 350~hPa, and any individual dry intrusion for which the tropopause lowering does not reach below 350~hPa is not considered deep enough for any irreversible mixing of $\mathrm{O}_{3}$ to have occurred during the STT \cite{Jaeg17}.
We first calculate the vertical integral of $\mathrm{O}_{3}$ from the maximum depth of the dry intrusion (PV$\geq$2 contour) up to 350~hPa to obtain the flux (kg~m$^{-2}$). For the average daily $\mathrm{O}_{3}$ flux, we track the number of days over which we calculate each individual intrusion flux and compute the mean over that number of days.
We also calculate the total average $\mathrm{O}_{3}$ flux (Tg~day$^{-1}$) in this manner, first as the vertical integral, then as the two dimensional integral of the positive column integrated $\mathrm{O}_{3}$ over latitude and longitude for each intrusion, and then average over the total number of days.

\subsection{Generating Composites}
\label{comp_ana}

To quantify the relative position and maximum of the average magnitude of the positive $\mathrm{O}_{3}$ anomalies near the AR events, statistically, we generate peak IVT-centric composites using the catalogued positions of the peak IVT for all AR time steps in December between 2004 and 2014.
Because the location of the $\mathrm{O}_{3}$ anomalies associated with each AR is dependent upon the AR direction of travel, we lessen the averaging out of anomalous $\mathrm{O}_{3}$ peaks and troughs by dividing all of the AR time steps into seven bins according to the direction of the peak IVT ($\Theta_{IVT}$).
We bin according to the direction in each tracked time step rather than individual AR events because ARs can change direction during their lifetime, enough that the relative locations of the dry intrusions are affected.
For each time step in a $\Theta_{IVT}$ bin, we define a 20$^\circ$ latitude x 20$^\circ$ longitude area centered on the peak IVTs and compute the mean IVT over that area for each $\Theta_{IVT}$ bin.
Over the same area, we compute the mean $\mathrm{O}_{3}$ anomaly for each $\Theta_{IVT}$ bin and the 200, 300, and 400~hPa atmospheric pressure levels.
We also compute the vertical cross-section of the composites in the same manner as for individual events, and determine the relative location of the maximum average $\mathrm{O}_{3}$ anomaly in these directionally divided composites.
Although averaging over a large range in $\Theta_{IVT}$ can dilute some of the synoptic scale information, the number of $\Theta_{IVT}$ bins was chosen primarily such that we can further divide those into bins of peak IVT and each bin would still contain adequate numbers to perform a statistical assessment.

We further examine the relationship between the composite maximum anomalous $\mathrm{O}_{3}$ and peak IVT by binning each of the directionally divided composites into two bins of low and high IVT.
The low and high bins of IVT are determined as the first (lowest) and fourth (highest) quartiles for all values of peak IVT tracked by our algorithm.
Each individual time step for all of the ARs that we track is binned according to these definitions in Section~\ref{o3vivt}.

\section{Case studies}
\label{casestudies}

To investigate the relationship between atmospheric rivers (ARs) and $\mathrm{O}_{3}$ anomalies within dry intrusions following the ARs, we take a two fold approach: 1) In-depth assessment of two case studies and 2) Statistical analysis via composites (Section~\ref{composite_results}).
The first approach is an exploration of the time series of AR events in order to first validate the use of MERRA-2 data using $\mathrm{O}_{3}$ profiles from OMI in November 2006 during an exceptional (Cat 5) AR event, and then to study the detailed time evolution of a series of strong, extreme, and exceptional (Cat 5) AR events that made landfall in December of 2010.
In this section, we begin to demonstrate the ubiquity of anomalous tropospheric $\mathrm{O}_{3}$ in dry intrusions associated with ARs, as well as detail the time evolution of both IVT and $\mathrm{O}_{3}$ concentration associated with single events.
This section demonstrates our algorithm and shows the similarity in the dynamics of individual, exceptional (Cat 5) AR events (06-07 November 2006 and 10-12 December 2010), while also contrasting the range of peak IVT values and $\mathrm{O}_{3}$ anomalies associated with a variety of AR events in December 2010.

\subsection{November 2006}
\label{2006}

\begin{figure}[!t]
 \centering
 \includegraphics[width=6.5in]{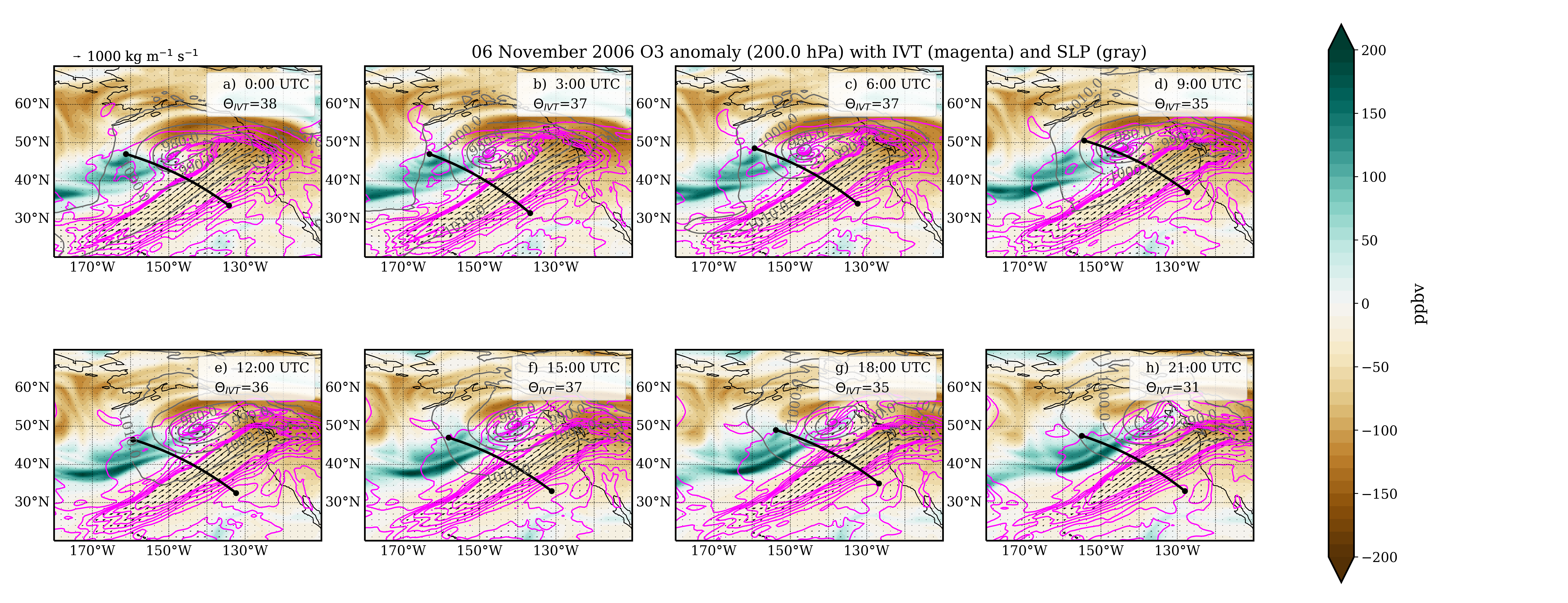}
 \caption{Anomalous $\mathrm{O}_{3}$ at 200~hPa (color scale) with AR IVT contours (magenta, steps of 100~kg~m$^{-1}$~s$^{-1}$, with the innermost contour 700~kg~m$^{-1}$~s$^{-1}$) and SLP contours (gray, 970-1010~hPa in steps of 10) in three-hour time steps on 06 November 2006. The arrows are an additional indicator of IVT amplitude and direction, and the scale is represented above the upper left plot. The legend reports the time and $\Theta_{IVT}$ at the location of the peak IVT for the AR at each time step. All of these data are from or computed from MERRA-2. The black line indicates the location of the vertical cross-section used for Figure~\ref{Nov06cross}, and stretches approximately across the location of the peak IVT in each time step.}
 \label{Nov06}
\end{figure}

\begin{figure}[!ht]
 \centering
 \includegraphics[width=6.5in]{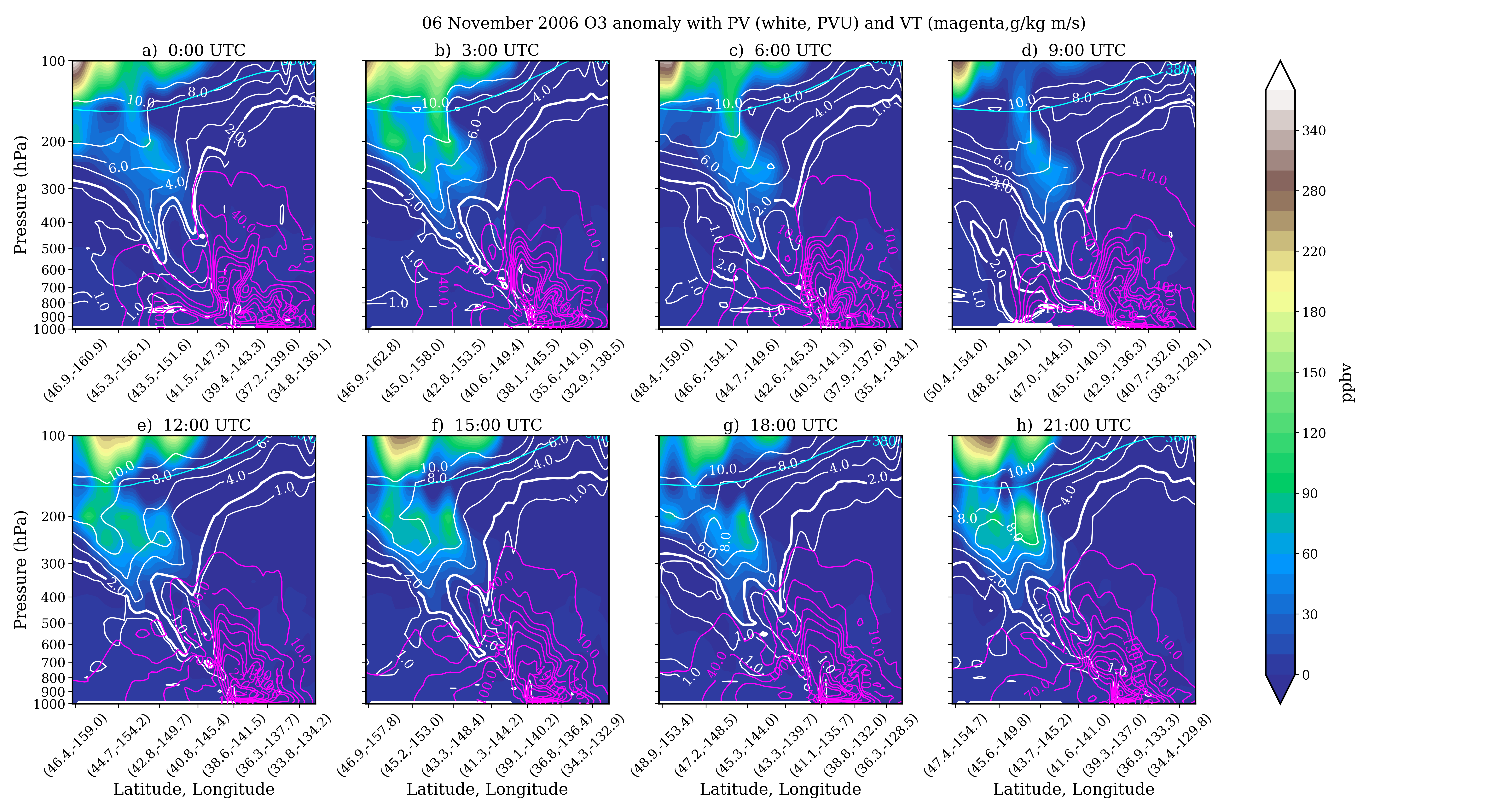}
 \caption{Vertical cross-section of positive anomalous $\mathrm{O}_{3}$ (color scale), PV contours (white; PV=2 contour bold), 380~K isotropic surface (cyan), and water vapor transport (VT, magenta) for each time step on 06 November 2006 as indicated by the black line in Figure~\ref{Nov06}, crossing approximately over the peak IVT of the AR for each time step. VT contours step from 10 to 340~g~kg$^{-1}$~m~s$^{-1}$ in steps of 30. All of these data are from or computed from MERRA-2. }
 \label{Nov06cross}
\end{figure}

On 06-07 November 2006 an exceptional (Cat 5, with maximum IVT exceeding 1300~kg~m$^{-1}$~s$^{-1}$) AR made landfall on the North American west coast, causing detrimental flooding and landslides.
An event of this scale occurs roughly once per year along the North American west coast \cite{Ralp19}, and we refer the reader to \citeA{Neim08a} for a detailed event description.
The initial discovery of the filamentary, positive $\mathrm{O}_{3}$ anomaly associated with ARs was in connection with this November 2006 event and is described in Section 4 of \citeA{Wang19}.
The average OMI data over the days 06-07 November 2006 produces anomalous total column water vapor $>$15~mm over the climatology. At a pressure level of 200~hPa, \citeA{Wang19} found a positive $\mathrm{O}_{3}$ anomaly filament with mixing ratios in excess of 120~ppbv positioned to the northwest and parallel to the AR. We show this $\mathrm{O}_{3}$ anomaly in the OMI data (Figure \ref{omi}, bottom row) and in MERRA-2 data (Figure \ref{omi}, top row) and plot it alongside the MERRA-2 IVT contours of the AR (Figure \ref{omi}, pink contours).
Compared to the OMI data, regridded to the MERRA-2 resolution of 0.5$^\circ$ x 0.625$^\circ$, we examine the MERRA-2 3 hourly snapshot of anomalous $\mathrm{O}_{3}$ and IVT for 06 and 07 November 2006 at 21:00 UTC (Figure~\ref{omi}, top row).
The OMI spectrometer on the Aura spacecraft crosses the equator at 13:30 LST, which is approximately 20:30 UTC over the NE Pacific, thus we compare to just the 21:00 UTC MERRA-2 3-hourly instantaneous data.

Overall, the $\mathrm{O}_{3}$ anomalies from OMI and MERRA-2 agree reasonably well with one another, showing the filamentary excess $\mathrm{O}_{3}$ following the AR IVT poleward and westward, with a mean bias of MERRA-2-OMI of 5.7 (42.5)\% and a mean absolute error of 14.5 (14.6)~ppbv for 06 (07) November 2006 at 200~hPa and the area shown in Figure~\ref{omi}.
Some discrepancies in the detailed structure and amplitude may be expected due to the limited vertical sensitivity of the OMI profiles, as well as the simplified chemistry in computing the MERRA-2 tropospheric $\mathrm{O}_{3}$.
However, at 200~hPa, which is close to the tropopause on average in this region of interest, 14.4\% and 12\% of the $\mathrm{O}_{3}$ is tropospheric for 06 and 07 November, respectively, and \citeA{Warg15} show that this simplified chemistry is sufficient for the assimilated upper troposphere and lower stratosphere (UT/LS) data product.
At this pressure level and these latitudes, the MERRA-2 $\mathrm{O}_{3}$ is found to have a mean bias of 1.2\% compared to ozonesondes without much seasonal dependence \cite{Warg15}, and the winter OMI-ozonesondes mean biases are comparable \cite{Huan17}.
Making the same comparison for the standard deviations of the differences, the UT/LS OMI-ozonesondes is between 18-27\% \cite{Huan17}, while similarly for MERRA2-ozonesondes these values for the UT/LS are 24.5\% and 11.2\%, respectively \cite{Warg17}.
Even so, at 200~hPa, this agreement (in addition to the aforementioned MERRA-2 validations) demonstrates the reliability of MERRA-2 $\mathrm{O}_{3}$ anomalies in the UT/LS, and the use of these data for identifying UT/LS $\mathrm{O}_{3}$ anomalies in connection with ARs and the subsequent analysis of this work.

In a similar comparison, \citeA{Jaeg17} examine MERRA-2 and MLS composite data of $\mathrm{O}_{3}$ concentration and anomalous $\mathrm{O}_{3}$ at 261~hPa centered on ECs, and they find these data to agree well. The assimilation of MLS $\mathrm{O}_{3}$ profiles in addition to the OMI total column $\mathrm{O}_{3}$ is an additional advantage to using the MERRA-2 data in this analysis.

We further investigate the time series data for this AR event using the $\mathrm{O}_{3}$ mixing ratios and IVT derived from MERRA-2.
Beginning on 01 November 2006, our algorithm follows the evolution of nine atmospheric river events throughout the entire month by tracking the peak value of IVT at each 3-hour time step as outlined in Section~\ref{IVTana}.
The algorithm begins tracking the early formation of this particular AR event on 04 November 2006 at 170$^\circ$~W and 32.5$^\circ$~N, and follows it until it completely diminishes on 08 November 2006, after making landfall on 06 November.
We simultaneously track the peak $\mathrm{O}_{3}$ anomaly as described in Section~\ref{o3ana}, as well as the separation between peak $\mathrm{O}_{3}$ anomaly and peak IVT.
Figure~\ref{Nov06} shows shaded maps of the anomalous $\mathrm{O}_{3}$ at 200~hPa in three-hourly time steps on 06 November 2006, with magenta IVT contours and gray sea level pressure (SLP$\leq$1010~hPa) contours overlaid.
This AR occurred in conjunction with an EC that can be identified via the low SLP.
On this day, the peak $\mathrm{O}_{3}$ values tracked by our algorithm are in excess of 120~ppbv for six out of the eight three-hour time steps, and the maximum excess for the entire day $\mathrm{O}_{3}$ exceeds 240~ppbv.

The thick, black line with two endpoints on each panel of Figure~\ref{Nov06} approximately crosses through the location of the maximum $\mathrm{O}_{3}$ anomaly and IVT, and is used to generate the vertical cross sections plotted in Figure~\ref{Nov06cross}.
These cross-sections highlight the vertical structure of anomalous $\mathrm{O}_{3}$, PV (white), and water vapor transport (VT, magenta).
The positive $\mathrm{O}_{3}$ anomalies follow the PV contours indicative of tropopause folding upstream, poleward, and westward of the high water vapor flux of the AR in each time step.
On 06 November, the dry intrusion bringing ozone-rich air reaches a pressure of at least 667~hPa as traced by these particular vertical cross sections (black lines in Figure~\ref{Nov06cross}).
Using the \citeA{Sker15} characterization of tropopause fold, this dry intrusion is classified as a medium fold with $\Delta$P~$>$~250~hPa.
The tropopause remains lowered throughout the duration of the event and well after landfall.
On 08 November 2006, two days after the AR  made landfall, and while it was still detectable, the positive $\mathrm{O}_{3}$ anomaly associated with the dry intrusion persisted with a significantly lowered dynamical tropopause at $p=$480~hPa.

\subsection{December 2010}
\label{2010}

\begin{figure}[!tbp]
 \centering
 \includegraphics[width=6.5in]{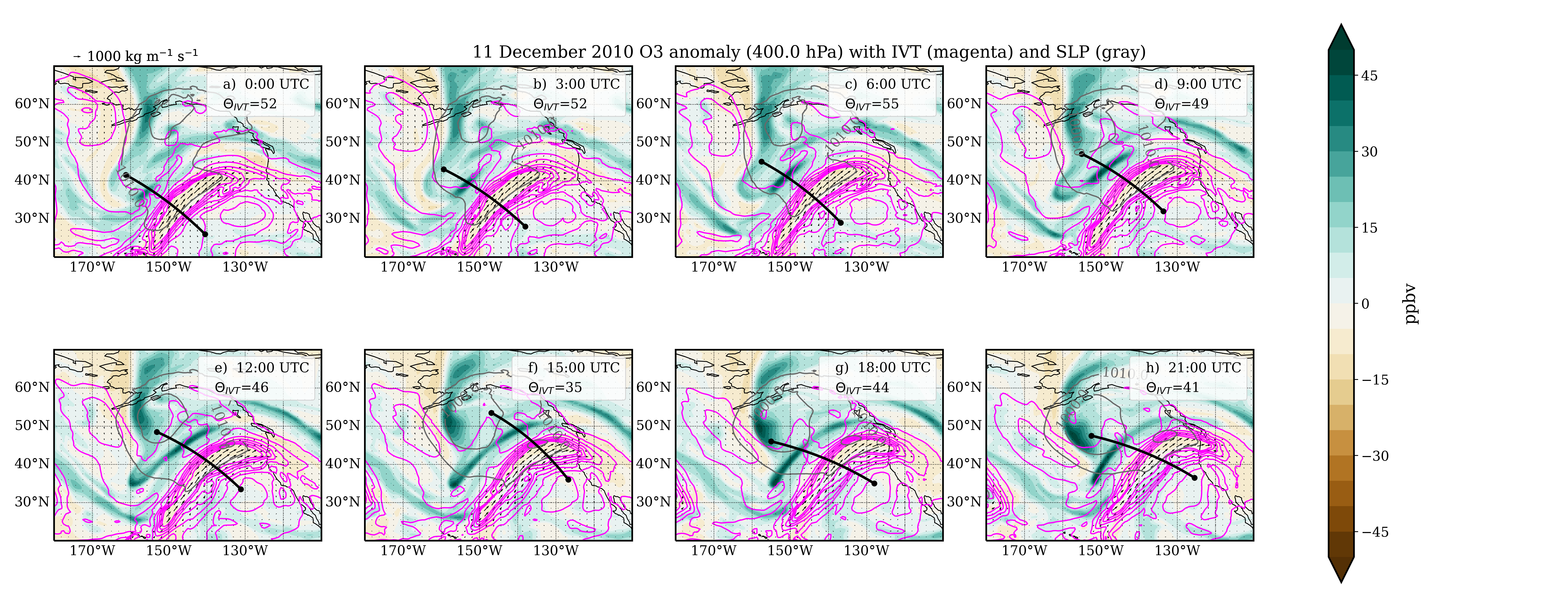}
 \caption{Anomalous $\mathrm{O}_{3}$ at 400~hPa (color scale) with AR IVT contours (magenta, steps of 100~kg~m$^{-1}$~s$^{-1}$, with the innermost contour 700~kg~m$^{-1}$~s$^{-1}$) and SLP contours (gray, 970-1010~hPa in steps of 10) in three hour time steps on 11 December 2010. The arrows are an additional indicator of IVT amplitude (length of arrow) and AR direction of travel, and the scale is represented above the upper left plot. The legend reports the time and $\Theta_{IVT}$ at the location of the peak IVT for the AR at each time step. All of these data are from or computed from MERRA-2. The black line indicates the location of the vertical cross-section used for Figure~\ref{Dec1110cross}, and stretches approximately across the location of the peak IVT in each time step.}
 \label{Dec1110}
\end{figure}

\begin{figure}[!tbp]
 \centering
 \includegraphics[width=6.5in]{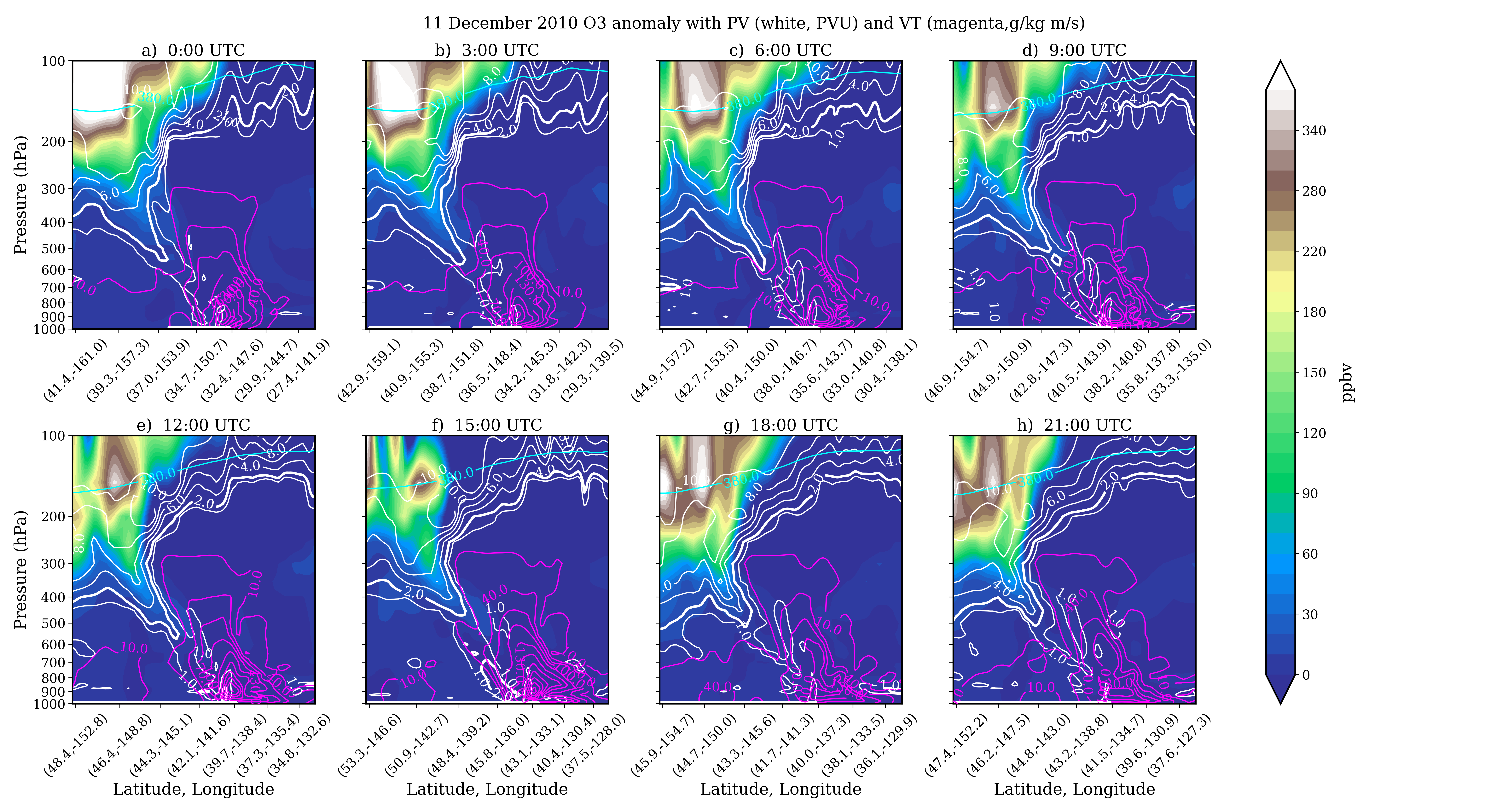}
 \caption{Vertical cross-section of positive anomalous $\mathrm{O}_{3}$ (color scale), PV contours (PVU; white, PV=2 bold), 380~K isotropic surface (cyan), and water vapor transport (VT, magenta) for the black line indicated in Figure~\ref{Dec1110}, crossing approximately over the peak IVT of the AR for each time step. VT contours step from 10 to 340~g~kg$^{-1}$~m~s$^{-1}$ in steps of 30. All of these data are from or computed from MERRA-2.}
 \label{Dec1110cross}
\end{figure}

\begin{figure}
    \centering
    \includegraphics[width=6.0in]{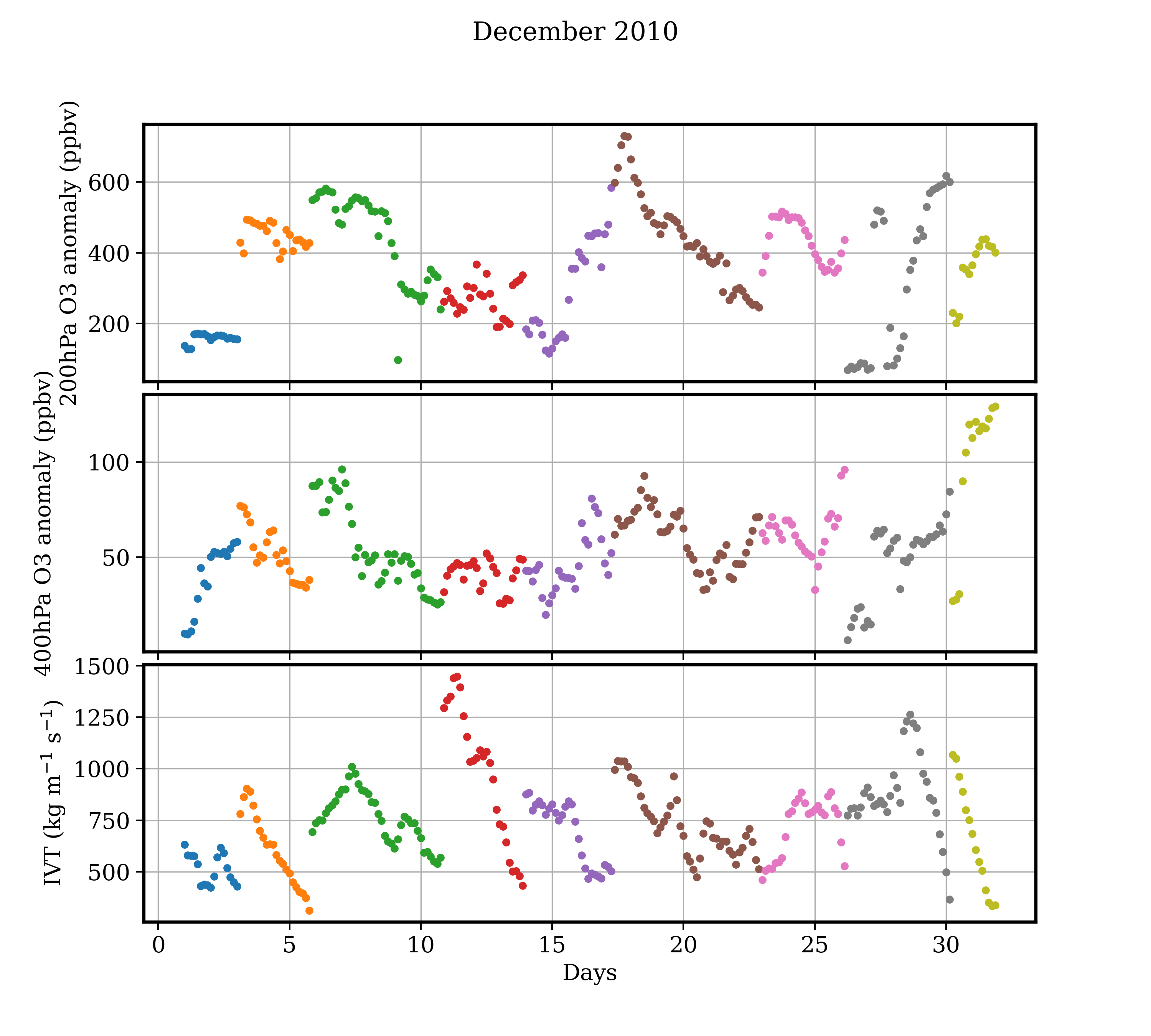}
    \caption{Time series of peak IVT (kg m$^{-1}$ s$^{-1}$, bottom) and peak $\mathrm{O}_{3}$ anomaly (ppbv) at 400~hPa (middle) and 200~hPa (top) within $\lesssim$2000~km poleward of the IVT location from MERRA-2 at each 3-hourly time step in December 2010. Our algorithm tracks the maximum IVT and associated peak $\mathrm{O}_{3}$ anomaly for each AR identified in the Rutz catalog at each time step until the AR diminishes, then searches for the next maximum AR event over the NE Pacific. Each color change indicates a new AR event as classified by our algorithm. There is no significant statistical correlation between IVT and peak $\mathrm{O}_{3}$ anomaly at 400 or 200~hPa for this time series.}
    \label{Dec2010peaks}
\end{figure}

As a second case study, we track all of the December 2010 AR events over the NE Pacific and quantify the associated dry intrusion and $\mathrm{O}_{3}$ anomalies correlated with them.
In this case, we also track the peak $\mathrm{O}_{3}$ anomalies on the 400~hPa pressure layer in addition to the 200~hPa level.
The 400~hPa pressure surface is insightful as the approximate height at which the water vapor transport intensity decreases to near zero and at which the positive anomalous $\mathrm{O}_{3}$ remains consistently significant above the climatology.
This pressure surface is also below the average level of the global extratropical dynamical tropopause, which is about 300~hPa, but varies significantly with latitude \cite{Sker14,Jaeg17}.
Over our region of interest and using the MERRA-2 data to determine the lower of the PV=2 or 380~K surface, we compute the average climatological dynamical tropopause to be 275~hPa.

ARs of December 2010 brought catastrophic consequences to the North American west coast.
\citeA{Ralp10} summarize the extreme precipitation due to two of these ARs' transport of massive amounts of water vapor from the Pacific Ocean to the North American west coast resulting in as much as 250-670~mm of rainfall in mountainous areas, causing flooding, and also providing crucial snow pack between 10-12 December and 20-22 December.
We track a total of nine ARs (by our algorithmic count) over the NE Pacific in December 2010, four of which were Cat 5 ARs (defined as having peak IVT$\ge$1000 and duration $\ge$48 hours or peak IVT$\ge$1250~kg~m$^{-1}$~s$^{-1}$ and duration$\ge$24 hours, \cite{Ralp19}) that made landfall on the North American west coast.

Figure~\ref{Dec1110} shows shaded maps of the anomalous $\mathrm{O}_{3}$ at 400~hPa with magenta IVT contours and gray SLP overlaid in three-hour time steps for 11 December 2010, a day in which the peak IVT reached $>$1400~kg~m$^{-1}$~s$^{-1}$.
As in Figure~\ref{Nov06}, the black geodesic line crosses approximately through the location of the peak $\mathrm{O}_{3}$ anomaly and IVT, and is used to generate the vertical cross sections shown in Figure~\ref{Dec1110cross}.
The peak anomalous $\mathrm{O}_{3}$ at 400~hPa varies between 20-50~ppbv over the course of this AR event, which formed on 10 December 2010, made landfall on 11 December, and dissipated on 13 December after dropping in IVT to below 500~kg~m$^{-1}$~s$^{-1}$ (Figure~\ref{Dec2010peaks},  red dots).
The vertical cross section (Figure~\ref{Dec1110cross}) reveals the tropopause fold via the PV contours (white) tilted eastward toward the AR, and it shows the positive anomalous $\mathrm{O}_{3}$ (shading) coincident with the fold.
At 200~hPa, the anomalous $\mathrm{O}_{3}$ reaches over 300~ppbv.
In Section~\ref{o3flux_discussion} below, we estimate and discuss the total $\mathrm{O}_{3}$ flux into the troposphere over the 4-day duration of this AR event.

There were several other ARs of strong and extreme intensity that made landfall in December 2010, so we plot in Figure~\ref{Dec2010peaks} the peak IVT and peak $\mathrm{O}_{3}$ anomaly at each 3-hour time step.
Each color indicates the tracking of an individual AR event until it dissipates, as defined by the binary tag catalog.
While some correlation between IVT and $\mathrm{O}_{3}$ anomaly peaks are discernible on short timescales, we find no statistically significant correlation in the time series across the entire month.
We also test a lag correlation with no statistical significance.
This lack of correlation may be due to spatial or temporal limitations of the algorithm.
It may be that the choice of tracking individual anomalous $\mathrm{O}_{3}$ points at 200~hPa and 400~hPa does not fully capture the dry intrusion depth.
It is also possible that the correlated relationship between anomalous $\mathrm{O}_{3}$ in dry intrusions following ARs is an average effect that depends on more global properties of the ARs, such as the entire duration of each event or its overall intensity.
In the following sections, we systematically explore the correlation between average AR IVT and anomalous $\mathrm{O}_{3}$ through examining composites binned according to the the magnitudes of their integrated vapor transport and quantifying the vertical structure of the anomalous $\mathrm{O}_{3}$, PV, and vapor transport of these composites.


\section{Statistical Summary and Composite Results}
\label{composite_results}

\begin{figure*}
\centering

    \includegraphics[width=4.5in]{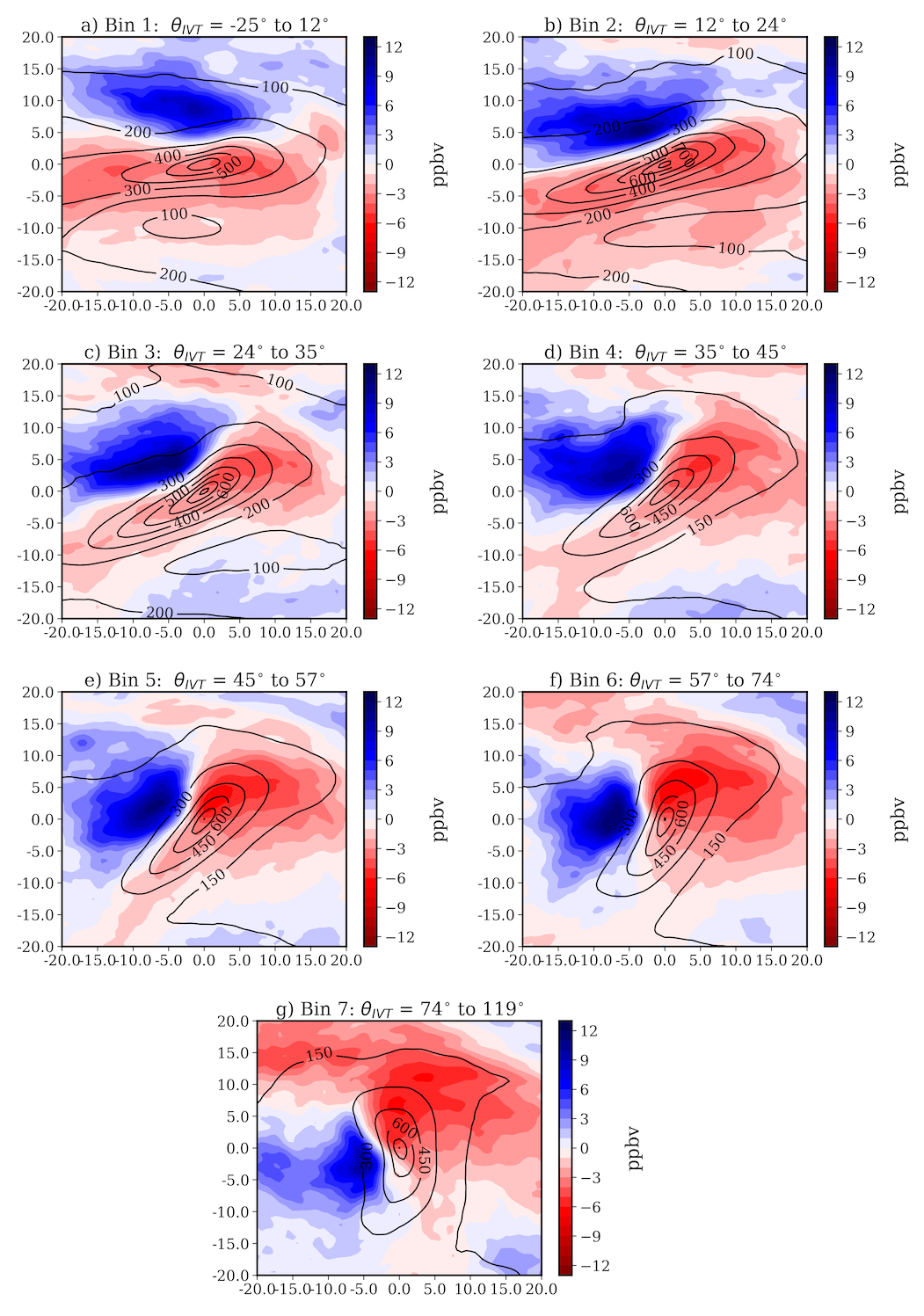}

\caption{Anomalous $\mathrm{O}_{3}$ at 400~hPa (shading) and AR IVT (black contours) composites for all December AR events between 2004-2014, divided by AR direction of travel, $\Theta_{IVT}$ into seven bins with N=382 AR time steps in each bin. Composites are generated from MERRA-2, and x and y axes are distance from the composite center in degrees longitude and degrees latitude, respectively, with x increasing in the east direction and y increasing in the poleward (north) direction.}
\label{composites1}
\end{figure*}

We now explore the relationship between ARs and $\mathrm{O}_{3}$ anomalies and dry intrusions statistically for all ARs occurring in the NE Pacific in December 2004-2014.
While ARs occur all year, their numbers are greatest over the NE Pacific (U.S. West Coast Landfalling) during the Northern Hemisphere cold months (November through February) \cite{Mund16}, so we focus on December.
Assessment of $\mathrm{O}_{3}$ anomalies associated with ARs indicates that the anomalies roughly trail behind the front of the AR, but the relative location depends on the direction of the AR (Figure~\ref{composites1}).
The morphology of the anomalies is often filamentary, like that of the ARs (e.g., Figure~\ref{omi}), but can also be more circular or clumped (e.g., Figure~\ref{Dec1110} top left).

\begin{figure*}
\centering

    \includegraphics[width=5.5in]{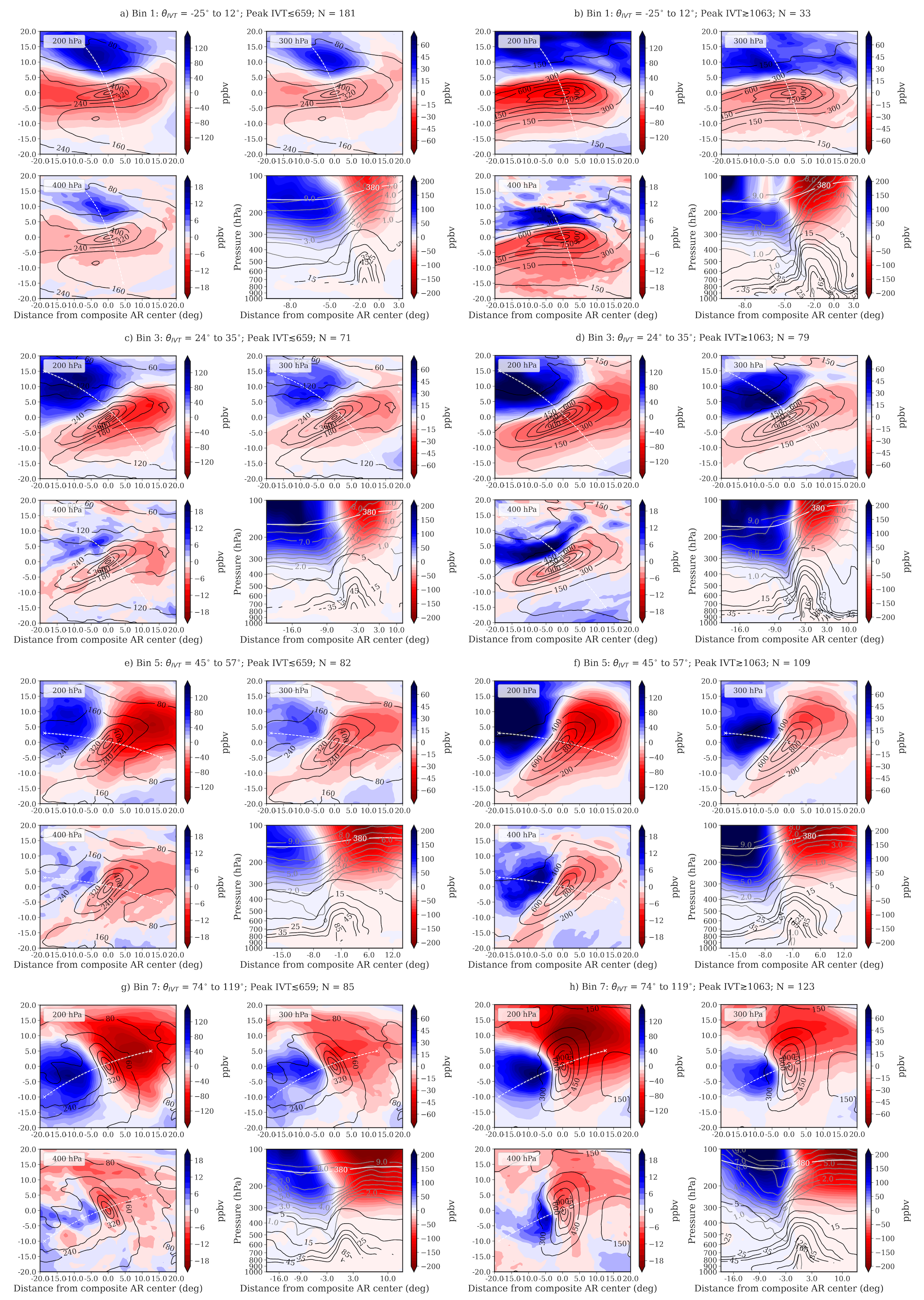}

\caption{Composite $\mathrm{O}_{3}$ anomalies (shading) and AR data (black contours) for bins of peak IVT$\lesssim$ 659~kg~m$^{-1}$~s$^{-1}$ (a, c, e, g) and  peak IVT$\gtrsim$ 1063~kg~m$^{-1}$~s$^{-1}$ (b, d, f, h). Rows of 2x2 plots are for bins of $\Theta_{IVT}$; we include here only bins 1, 3, 5, and 7, and bins 2, 4, and 6 are in the appendix (Figure~\ref{composites3}). Composites generated from MERRA-2. Each set of 2x2 plots depicts the $\mathrm{O}_{3}$ anomalies at 200~hPa (upper left), 300~hPa (upper right), and 400~hPa (lower left), as well as the vertical cross-section defined by the white dashed line over pressure levels 1000-100~hPa (lower right). These figures exemplify the composite increased $\mathrm{O}_{3}$ anomaly concentration as a function of increased IVT.}
\label{composites2}
\end{figure*}

Figure \ref{composites1} shows composite (average) $\mathrm{O}_{3}$ anomaly (shading) at 400~hPa and IVT (black contours) for all three hour time steps in December 2004-2014 centered on the AR peak IVT for each time step and divided into seven bins of AR direction, as described in Section~\ref{comp_ana}.
The composites are oriented with x increasing in the east direction and y increasing in the poleward (north) direction, but the axes depict distance from the AR center (peak IVT location) in degrees.
While many of the details of individual events have been averaged out due to latitudinal and longitudinal differences in the exact location of the anomalies relative to each AR center, the composites demonstrate the general direction of the $\mathrm{O}_{3}$ anomalies associated with AR events.
The positive anomalous $\mathrm{O}_{3}$ occurs along the ARs' long axis to the north/west, depending on the IVT angle, while negative $\mathrm{O}_{3}$ anomalies are observed within the AR region (Figure~\ref{composites1}), potentially as a result of troposphere-stratosphere transport (TST, see Figure~\ref{composites2} for more vertical information).
The location of the peak value of $\mathrm{O}_{3}$ anomaly relative to ARs traveling east/northeast is roughly north/northwest (Figure~\ref{composites1}a-e), with the possibility of the maximum anomalous $\mathrm{O}_{3}$ being a few degrees eastward of the peak IVT value if the direction of the IVT is east/southeast (ARs included in Figure~\ref{composites1}a).
For an AR traveling north or northeast, the $\mathrm{O}_{3}$ anomaly lies directly west or southwest of the AR center (Figure~\ref{composites1}e-g).
We analyze only the positive $\mathrm{O}_{3}$ anomalies that result from STT, and do not interpret the negative anomalies over the AR region due to the greater uncertainty of the MERRA-2 data in the lower troposphere.
The STT is reliable as MERRA-2 adequately captures the upper tropospheric and lower stratospheric $\mathrm{O}_{3}$ concentrations \cite{Know17}.

We see that the composites at 400~hPa have average peak $\mathrm{O}_{3}$ anomalies ranging from 10-13~ppbv for all AR travel directions.
For 95\% of the individual AR time steps, the peak $\mathrm{O}_{3}$ anomaly occurs between 500-1600 km from the center (or peak IVT) of the AR.
In the composites, we find that the peak average $\mathrm{O}_{3}$ anomaly occurs roughly 700-1000~km from the AR center, independent of the AR direction of travel.
The only difference is the relative direction of the $\mathrm{O}_{3}$ anomaly driven by the folding or lowering of the tropopause along the poleward or westward length of the ARs.
While we use a different definition of $\mathrm{O}_{3}$ anomaly than \citeA{Jaeg17}, these anomalous $\mathrm{O}_{3}$ amplitudes are commensurate in what they find for cyclone composites at the same pressure level.

\begin{figure}[!tbp]
 \centering
 \includegraphics[width=6.0in]{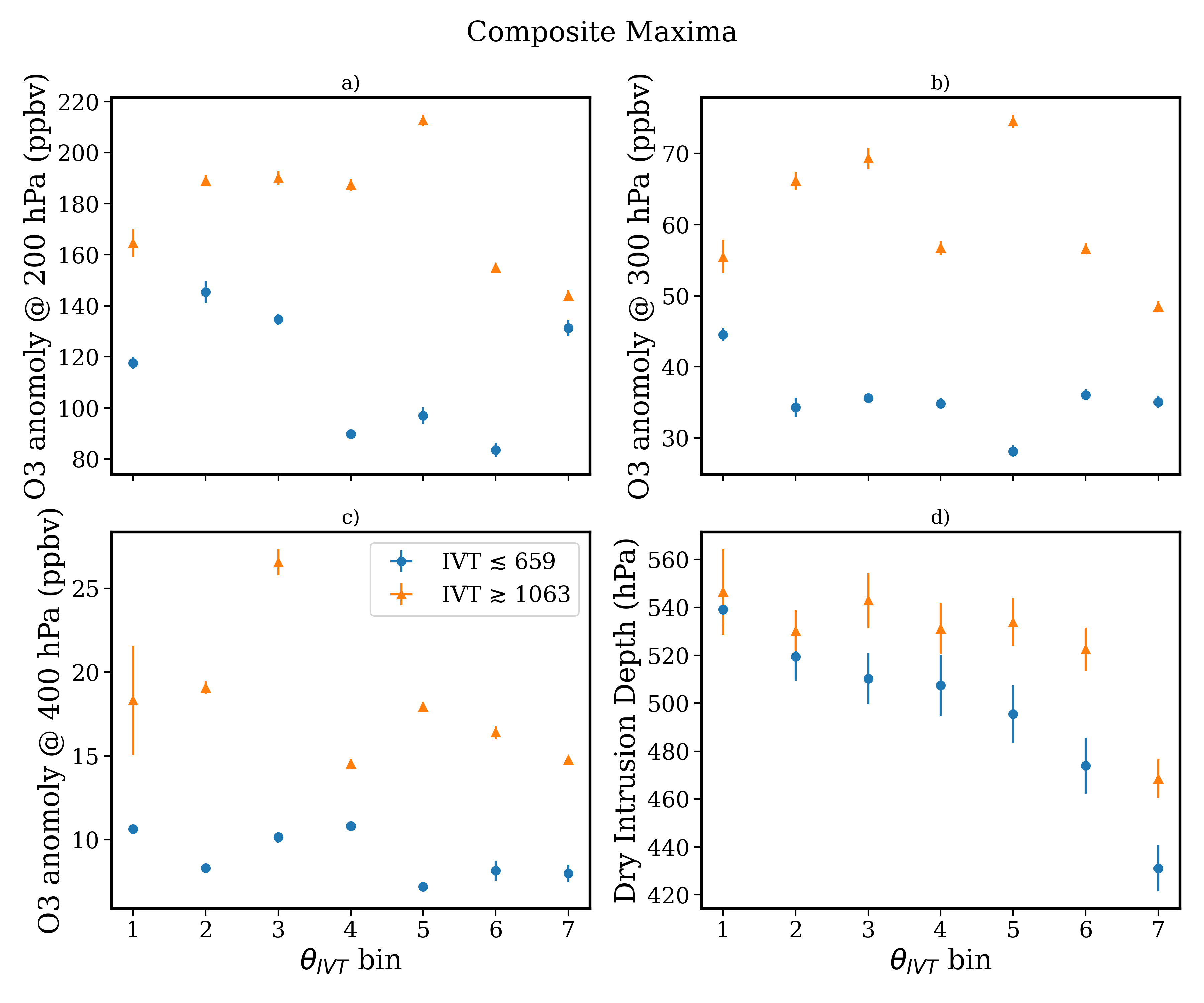}
 \caption{Composite maximum $\mathrm{O}_{3}$ anomaly concentration for each $\Theta_{IVT}$ bin divided further by peak IVT (blue circles: peak IVT$\lesssim$659~kg~m$^{-1}$~s$^{-1}$; orange triangles: peak IVT$\gtrsim$1063~kg~m$^{-1}$~s$^{-1}$) for latitude by longitude cross sections at 200 (upper left), 300 (upper right), and 400~hPa (lower left). The uncertainties are the standard deviation of all means from 100 bootstrap realizations of the composites. The lower right plot shows the average depth of the dry intrusion as indicated by the PV=2 contour for all AR vertical cross-sections in a given bin. These uncertainties reflect the standard error on the mean.  All values are derived from MERRA-2.}
 \label{peakO3anom_max2PV}
\end{figure}

\subsection{Composite $\mathrm{O}_{3}$ anomaly concentration as a function of AR IVT}
\label{o3vivt}

We assess any correlation between the amplitude of positive $\mathrm{O}_{3}$ anomaly and intensity of AR by further binning the ARs into two bins of peak IVT while still binning by direction of travel for the reasons described in the previous section.
Using all of the AR IVTs across all directional bins, we define two bins: ``low" for values of peak IVT in the first quartile and ``high" for values of peak IVT in the fourth quartile of all event times that we track.
These low and high bins correspond to all peak IVT less than 659~kg~m$^{-1}$~s$^{-1}$ and greater than 1063~kg~m$^{-1}$~s$^{-1}$, respectively.
As a result, there are not necessarily the same number of AR IVTs in each bin because each $\Theta_{IVT}$ bin may have a different IVT quartile ranges compared to the combined data set of all directions of ARs.
These bins of peak IVT roughly separate the events into weak (Cat 1) and moderate (Cat 2) AR IVT values in the low bin and extreme (Cat 4) and exceptional (Cat 5) in the high bins, although in this part of the analysis we do not take into account duration nor the absolute maximum value of IVT over an entire event \cite<for full categorization definitions see>{Ralp19}.
That is, time steps of a given AR event for which peak IVT is in the first quartile will be in the low bin and time steps of that same AR for which peak IVT is in the fourth quartile (if present) will be in the high bin.

Figures~\ref{composites2} and \ref{composites3} depict the three-dimensional structure of the $\mathrm{O}_{3}$ anomalies associated with atmospheric rivers divided into bins by AR direction of travel and peak IVT values.
Each two-by-two grid in the figures corresponds to the average anomalous $\mathrm{O}_{3}$ at 200 (top left), 300 (top right), and 400~hPa (bottom left) pressure levels in latitude by longitude plotted as a function of distance from the center of AR peak IVT values.
The vertical cross-section (bottom right) is computed across the white dashed line in the other three figures and spanning pressure levels 1000-100~hPa.
Each of the two-row blocks in the figures corresponds to the seven bins of AR direction of travel, and the two columns of two by two blocks correspond to the two bins of peak IVT value (left: peak IVT$\lesssim$ 659~kg~m$^{-1}$~s$^{-1}$; right:  peak IVT$\gtrsim$ 1059~kg~m$^{-1}$~s$^{-1}$).
The latitude by longitude cutouts at 200, 300, and 400~hPa help to demonstrate the dry intrusion's breadth and the filamentary shape at 400~hPa along the length of the AR.
The vertical (pressure level) cross-section (bottom right of each 2x2 grid, a-h) indicates the average depth of the dry intrusions and the west-to-east directionality as seen via the $\mathrm{O}_{3}$ anomaly and the PV contours.
The cross section is always shown from the approximate location of the maximum $\mathrm{O}_{3}$ anomaly across the width of the AR. The PV intrusion is on average tilted toward the AR, regardless of AR direction of travel; we also saw this clearly in the case study events and time step cross sections in Figures~\ref{Nov06cross} and \ref{Dec1110cross}.
Once again, individual details of the lat~x~lon anomalies, and even more so, the depth of the dry intrusion, are averaged out by the slight differences in location of the intrusions and peak $\mathrm{O}_{3}$ values relative to the center (peak IVT) of the AR at each time step.
Here (Figure~\ref{composites2}) we include only $\Theta_{IVT}$ bins 1, 3, 5, and 7 as examples, and we placed the rest of the corresponding figures for bins 2, 4, and 6 in the appendix (Figure~\ref{composites3}).
Each $\Theta_{IVT}$ bin tells a similar story regarding the division with respect to the two peak IVT bins, and more details are provided in Figure~\ref{peakO3anom_max2PV} for all of the directional bins.

In Figure~\ref{peakO3anom_max2PV}, the difference in magnitude of the $\mathrm{O}_{3}$ anomalies in the low versus high peak IVT bins is clear: the low peak IVT bins correspond to smaller maximum $\mathrm{O}_{3}$ anomalies in the composites at all pressure levels.
At 400~hPa, the maximum value of the composite anomalous $\mathrm{O}_{3}$ ranges between 14.5$\pm$0.3 to 25.5$\pm$0.8~ppbv for the peak IVT~$\gtrsim$1063~kg~m$^{-1}$~s$^{-1}$ bin, and 7.1$\pm$0.3 to 10.8$\pm$0.3~ppbv in the low IVT bin.
The uncertainties are the standard deviation of all means from 100 bootstrap realizations of the composites.
Through this statistical analysis we observe that there is in fact a relationship between anomalous $\mathrm{O}_{3}$ and AR intensity.

We investigate whether the anomalous $\mathrm{O}_{3}$ concentrations are driven by the depth of the dry intrusions by tracking the maximum pressure reached by the PV=2 contour within the dry intrusion for each time step of each AR (see Section~\ref{o3ana} for intrusion identification).
In the lower right panel of Figure~\ref{peakO3anom_max2PV}, we plot the average depth of the dry intrusions computed for the low versus high peak IVT bins. The uncertainties reflect the standard error on the mean.
We find that while the high IVT bin dry intrusions are on average deeper, there is not a clear statistical divide for all bins of $\Theta_{IVT}$.
A two-sided Kolmogorov–Smirnov test on the entire distributions of dry intrusion depths for all bins of $\Theta_{IVT}$ separated only into bins of low (first quartile) and high (fourth quartile) IVT produces a p-value of 0.025, rejecting the null hypothesis that the samples are drawn from the same distribution.
Performing this test on each $\Theta_{IVT}$ bin, we reject the null hypothesis for bins 3, 5, 6, and 7; however, this test does not account for the uncertainty in the bins.
Again, averaging according to peak IVT value over all 3 hour time steps may be detracting from the details of the individual events.
The downward trend in average depth indicates that tracing the dry intrusion along an eastward travelling AR yields, on average, deeper intrusion events; while an east-west cross section cutting through northward ARs results in a more shallow tracing of the dry intrusion.
This is likely explained by the orientation of the cold front with respect to AR direction of travel.

Another way to assess the relationship of ARs to the depth of associated dry intrusion events is in the distribution of depths in relation to all AR events.
Over the course of the eleven Decembers studied in this work, we track 138 AR events.
Of these, 39 of them are connected to a dry intrusion that reaches below 700~hPa in at least one of the time steps throughout its duration; this amounts to 28.3\% of all AR events we track.
Our algorithm for tracking the tropopause folding or lowering ensures the tracking of the PV=2 contour connected to the stratosphere and not part of a cutoff (Section~\ref{o3ana}), and we visually inspected this set of ARs to ensure the depth of the dry intrusions reach at least 700~hPa.
The percentage of the ARs connected to dry intrusions that are classified as deep (with $\Delta$P$\geq$350~hPa, \cite{Sker15}) in at least one time step for the duration of the AR event is 79.7\%.
If we compute the same statistic for all times in which the peak IVT is in the first quartile (low IVT bin), this includes a total of 116 ARs for which 7.7\% and 46.5\% of them are connected to dry intrusions that in at least one time step reach below 700~hPa and are classified as deep, respectively.
For the fourth quartile (high IVT bin) case, there are a total of 81 ARs for which the below 700~hPa and deep classifications are 16\% and 64.2\%.
This total AR event analysis indicates that a larger fraction of the high IVT bin ARs are connected to deeper dry intrusions and those which more frequently reach blow 700~hPa.

\subsection{Composite STT $\mathrm{O}_{3}$ flux as a function of AR IVT and in comparison to other work}
\label{o3flux_discussion}

\begin{figure}[!tbp]
 \centering
 \includegraphics[width=6.0in]{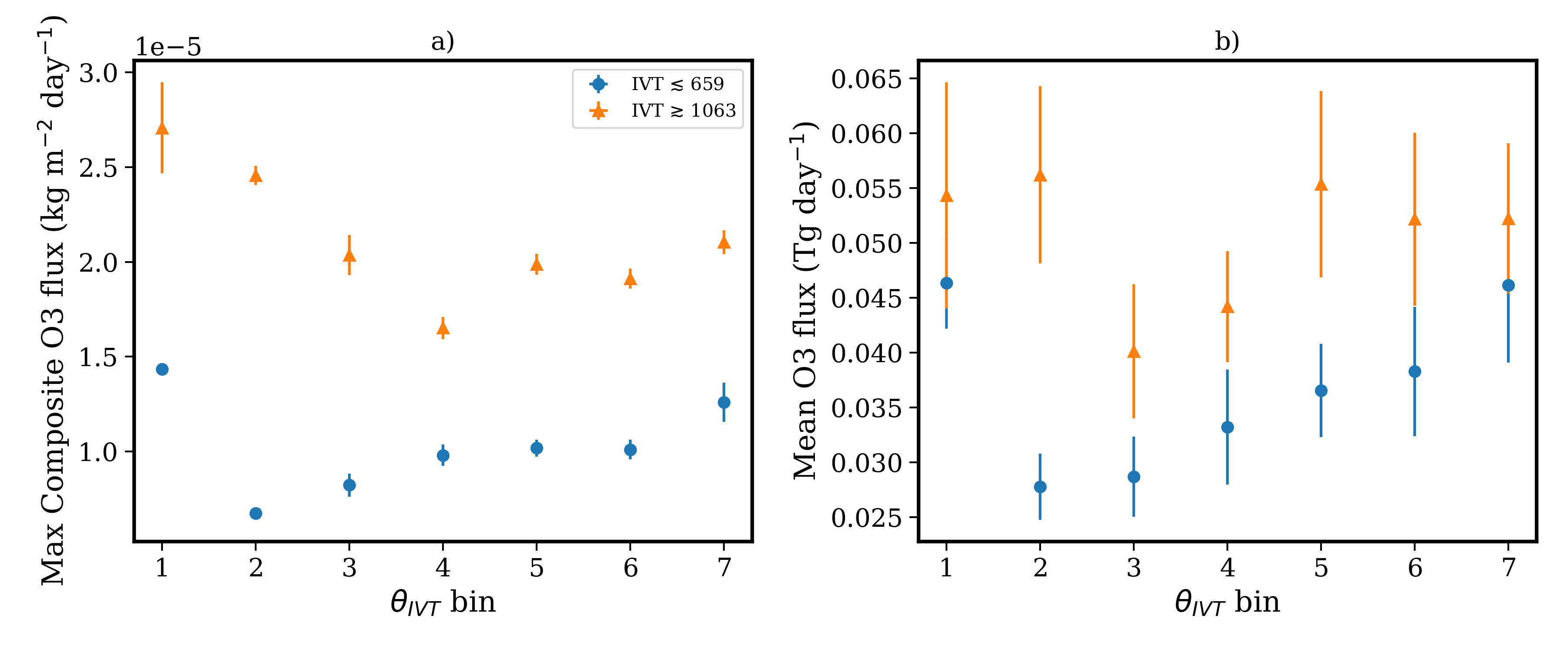}
 \caption{Left: The maximum value of STT $\mathrm{O}_{3}$ flux (kg~m$^{-2}$~day$^{-1}$) from the composites generated for each $\Theta_{IVT}$ bin (see Figures~\ref{composites1} and \ref{composites2} panel titles for angle ranges) for the first and fourth quartile peak IVT ARs (blue circles: peak IVT$\lesssim$659~kg~m$^{-1}$~s$^{-1}$; orange triangles: peak IVT$\gtrsim$1063~kg~m$^{-1}$~s$^{-1}$). The uncertainties are the standard deviation of the maximum of all means from 100 bootstrap realizations of the composites. Right: Average total area-integrated $\mathrm{O}_{3}$ flux (Tg~day$^{-1}$) for all events in each bin. Uncertainties reflect the standard error on the mean.  All values are derived from MERRA-2.}
 \label{o3flux}
\end{figure}

With the MERRA-2 data products, we quantify the total flux of $\mathrm{O}_{3}$ from the stratosphere into the troposphere (via STT) in relation to ARs following the methods described in Section~\ref{o3ana}.
We do not attempt to compute the TST since the MERRA-2 $\mathrm{O}_{3}$ data has larger bias in the upper troposphere as discussed in Section~\ref{data} \cite{Warg17}.
Assessing the percentage of $\mathrm{O}_{3}$ that reaches our constraining depth in the dry intrusion and is then pulled back up into the stratosphere is beyond the scope of this work, and thus, we advise careful consideration of these flux constraints.
For the purpose of this exercise of comparing total $\mathrm{O}_{3}$ flux in dry intrusion events associated with ARs to those from ECs, and other STT events more generally, we assume that the stratospheric $\mathrm{O}_{3}$ within the dry intrusion as defined by these PV limits is irreversibly mixed into the troposphere \cite{Jaeg17, Stoh03, Rood97, Lama94}.

Figure~\ref{o3flux}, left, shows the maximum composite $\mathrm{O}_{3}$ flux (left, kg~m$^{-2}$~day$^{-1}$) as a function of $\Theta_{IVT}$ for the high and low IVT bins, in orange and blue, respectively computed for all ARs that we track in December 2004-2014.
The uncertainties are the standard deviation of the maximum of all means from 100 bootstrap realizations of the composites.
Latitude~x~longitude composites of $\mathrm{O}_{3}$ flux centered on the AR peak IVT locations for dry intrusions with depth P$\gtrsim$350~hPa, divided into bins of direction and peak IVT, are plotted in the appendix, Figures~\ref{fluxcomposites1} and \ref{fluxcomposites2}.
The right panel of Figure~\ref{o3flux} shows the average total $\mathrm{O}_{3}$ flux (Tg~day$^{-1}$) integrated over the 20$^\circ$ lat x 20$^\circ$ lon area around the peak IVT for each time step of the ARs and averaged over the total number of days in each bin.
These area-integrated fluxes are computed individually for each time step of each AR in order to properly account for the distance covered with respect to latitude over the total area as the ARs move.
In the IVT$\lesssim$659~kg~m$^{-1}$~s$^{-1}$ bins, we find average $\mathrm{O}_{3}$ fluxes ranging from 0.028$\pm$0.003 to 0.046$\pm$0.007~Tg~day$^{-1}$, and in the IVT$\gtrsim$1063~kg~m$^{-1}$~s$^{-1}$ bins, the average fluxes range from 0.04$\pm$0.006 to 0.056$\pm$0.008~Tg~day$^{-1}$.
The uncertainties shown in the right-hand panel of Figure~\ref{o3flux} reflect the standard error on the mean, whereas the full range of area integrated fluxes for all ARs span 0.002-0.21~Tg~day$^{-1}$.
Considering all $\Theta_{IVT}$ bins, we do find a measurable difference in the total area integrated $\mathrm{O}_{3}$ flux for our two bins of IVT, with lower intensity events resulting in lower STT $\mathrm{O}_{3}$ flux.
We performed the same computation for the case study event on 10-12 December 2010, and found an average $\mathrm{O}_{3}$ flux of 0.15~Tg~day$^{-1}$.
All of these values are consistent with means and individual case studies for ECs:
\citeA{Jaeg17} find mean STT $\mathrm{O}_{3}$ flux from cyclones in the DJF season to be 0.075~Tg~day$^{-1}$, with total values ranging from 0.01 to 0.24~Tg~day$^{-1}$ and a strong dependence on 300~hPa wind speed;
Other EC-associated STT $\mathrm{O}_{3}$ flux measurements range from 0.01-0.15~Tg~day$^{-1}$ for a variety of individual events as well as averages computed via ground-based lidar measurements and Lagrangian trajectories \cite{Ance91,Lama94,Ebel96,Beek97,Coop04,Kuan12}.

While a complete assessment of NH STT $\mathrm{O}_{3}$ flux in connection to AR occurrences in contract to that in non-AR occurrences is beyond the scope of this paper, we can provide an estimation based on the others' climatological STT $\mathrm{O}_{3}$ flux as a function of geography and time, namely we use the results from \citeA{Sker14}.
\citeA{Sker14} Figure~18a shows the monthly STT $\mathrm{O}_{3}$ flux for the Northern Hemisphere averaged from 1979-2011.
From this figure, the December average is approximately 14$\pm$2~Tg~month$^{-1}$.
If we consider the average STT $\mathrm{O}_{3}$ flux from AR events that we track of 0.058$\pm$0.003~Tg~day$^{-1}$ and multiply by 31 days per month in December, this amounts to 1.80$\pm$0.09~Tg~month$^{-1}$, or between 13$\pm$2\% of the total December NH STT $\mathrm{O}_{3}$ flux.
Our method here only tracks one AR at a time through its lifetime over the NE Pacific.
In order to compare the AR dry intrusion contribution directly to the total STT $\mathrm{O}_{3}$ flux in the NH, we would need to track all ARs in the NH simultaneously -- a specific task that we reserve for future work.
We give a rough approximation by considering that 4-5 ARs occur simultaneously around the globe \cite{Ralp11}.
If half of these are active in the Northern Hemisphere, we approximate that the total STT $\mathrm{O}_{3}$ flux due to dry intrusions following ARs is 3.6-4.5~Tg~month$^{-1}$ for December; or as much as 32\% of the total NH STT $\mathrm{O}_{3}$ flux.

Furthermore, \citeA{Jaeg17} find similar mean $\mathrm{O}_{3}$ fluxes per cyclone to our average value, 0.057~Tg~day$^{-1}$.
Summing over all NH cyclones (4-6 per day), they reported daily mean $\mathrm{O}_{3}$ total STT flux from all NH cyclones to range from 0.1-0.5~Tg~day$^{-1}$ with a total contribution to the annual flux of 119~Tg~year$^{-1}$, which suggest the EC dry intrusion STT $\mathrm{O}_{3}$ flux makes up $\sim$42\% of all annual NH STT $\mathrm{O}_{3}$ flux \cite<281~Tg~year$^{-1}$,>{Jaeg17}.
Given the strong link between the occurrence of ECs and ARs, some fraction of the $\mathrm{O}_{3}$ flux in dry intrusions associated with each of these phenomena is overlapping; thus these are not independent results.
Though ARs coincide with an EC 82\% of the time in the North Pacific, AR location and duration cannot be determined from the EC \cite{zhan19}.
In the North Atlantic, for example, \citeA{Sode13} find that the lifetime of a single AR can span several ECs.
Thus, disentangling the STT $\mathrm{O}_{3}$ flux contribution in dry intrusions associated with ECs and ARs, respectively, is a complex task.

\section{Summary and Conclusion}
\label{conclude}
We have demonstrated for the first time positive $\mathrm{O}_{3}$ anomalies in the troposphere associated with atmospheric river events over the NE Pacific ocean for December 2004-2014.
These anomalies result from intrusions of dry, ozone-rich stratospheric air into the troposphere via STT following to the north/west of the AR's travel direction.
While ARs are associated with ECs (ECs) 82\% of the time, the exact location, duration, and intensity cannot be determined from the EC \cite{zhan19}.
We thus demonstrate the value in assessing the general location, duration, and magnitude of the $\mathrm{O}_{3}$ anomalies connected to the AR events.

Individual case studies of extreme AR events in November 2006 and December 2010 show STT of $\mathrm{O}_{3}$ in dry intrusions with excess $\mathrm{O}_{3}$ concentrations compared to the climatology reaching hundreds of parts per billion at 200~hPa near the tropopause, and consistently in excess of 50~ppbv in the upper troposphere at 400~hPa (Figure~\ref{Dec2010peaks}).
Composites for all AR events in December 2004-2014 show $\mathrm{O}_{3}$ anomalies in the upper troposphere within dry intrusions span up to the full average length of all ARs within a given $\Theta_{IVT}$ (direction of AR) bin.
Moreover, the average amplitude of excess $\mathrm{O}_{3}$ concentration at 400~hPa and total $\mathrm{O}_{3}$ flux in the dry intrusions increases with increasing AR IVT.
For all December AR events included in this analysis, the associated dry intrusion $\mathrm{O}_{3}$ flux per AR event ranges from 0.001 to 0.16~Tg~day$^{-1}$, with a median of 0.06~Tg~day$^{-1}$.
If we assume that the average flux is representative of all ARs across the globe at a given time, then this would indicate the dry intrusion STT $\mathrm{O}_{3}$ flux connected to ARs could be responsible for up to 32\% of the total NH STT $\mathrm{O}_{3}$ flux for December \cite{Sker14, Jaeg17}.

In light of the projected increase in both intensity and AR-day frequency of ARs with climate change \cite{Dett11b, Espi18, Huan20, Payn20, Rhoa21}, quantifying and closely monitoring dynamically-related phenomena such as this is critical.
Our work here implies that the increased intensity and/or frequency of ARs may also bring increased STT $\mathrm{O}_{3}$ flux in the warming future.
The overall levels of tropospheric $\mathrm{O}_{3}$ concentration as a result are dependent on a thorough assessment of the TST, and residence times need to be fully considered in conjunction.
The primary impact, however, of AR-associated dry intrusions is on the upper troposphere, which is highly consequential for the outgoing longwave radiation \cite{Bowm13,Doni15}. In fact, the longwave radiative effect may be most sensitive to $\mathrm{O}_{3}$ concentrations in the upper troposphere \cite{Laci90, Word11}, thus emphasizing the importance of quantifying UT/LS $\mathrm{O}_{3}$ concentrations and STT $\mathrm{O}_{3}$ flux from all sources.

\appendix

\section{Additional Composite Figures}
\begin{figure*}[!ht]
\centering

    \includegraphics[width=5.5in]{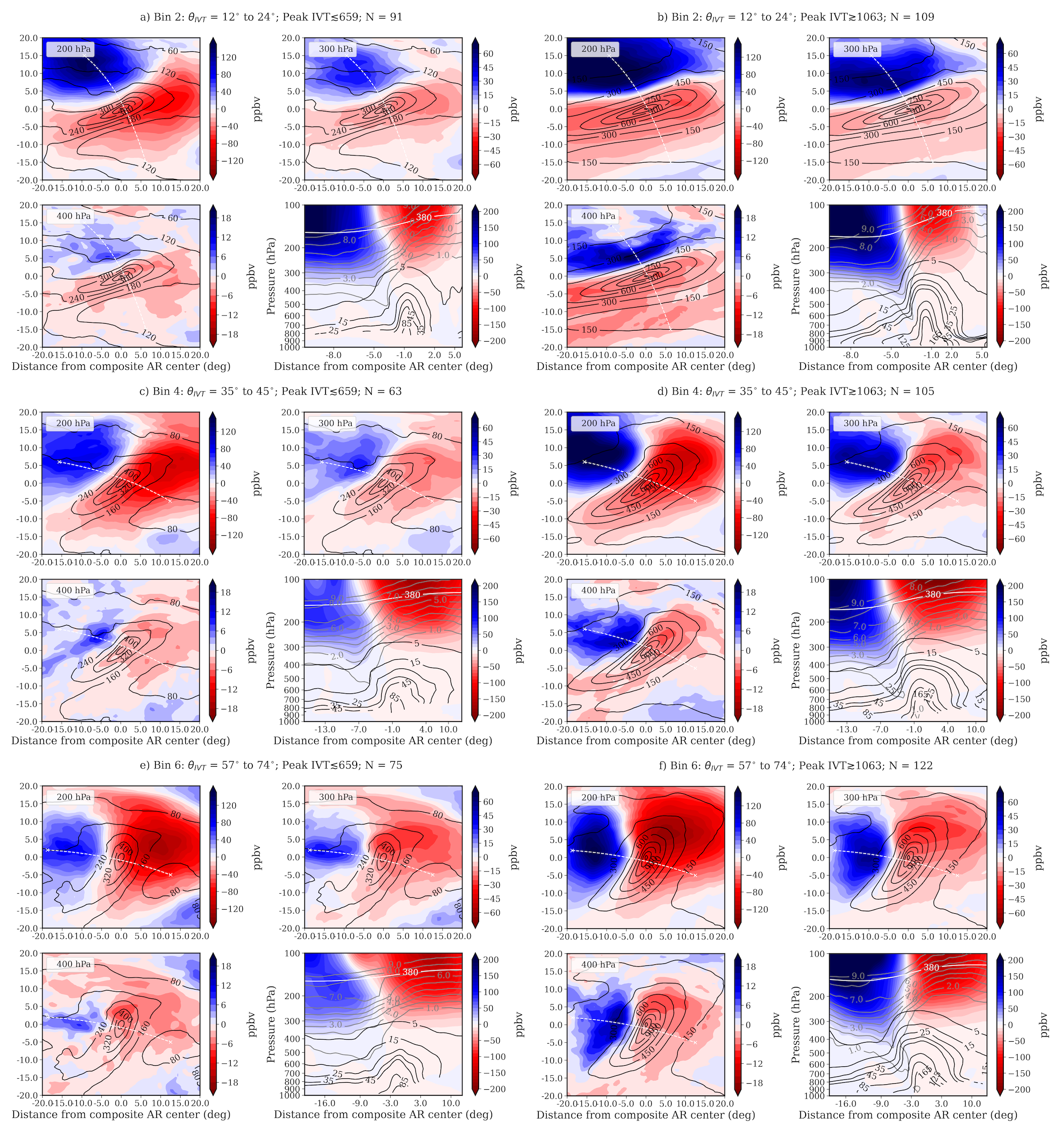}

\caption{Composite $\mathrm{O}_{3}$ anomalies (shading) and AR data (black contours) for bins of peak IVT$\lesssim$ 659~kg~m$^{-1}$~s$^{-1}$ (left column of 2x2 plots) and  peak IVT$\gtrsim$ 1063~kg~m$^{-1}$~s$^{-1}$ (right column of 2x2 plots). Composites generated from MERRA-2. Rows of 2x2 plots are for bins of $\Theta_{IVT}$; we include here bins 2, 4, and 6, as bins 1, 3, 5, and 7 are included in the main body (Figure~\ref{composites2}). Each set of 2x2 plots depict the $\mathrm{O}_{3}$ anomalies at 200~hPa (upper left), 300~hPa (upper right), and 400~hPa (lower left), as well as the vertical cross section defined by the white dashed line over pressure levels 1000-100~hPa (lower right). These figures exemplify the composite increased $\mathrm{O}_{3}$ anomaly concentration as a function of increased IVT. }
\label{composites3}
\end{figure*}

Here we include important, additional figures that are not vital for the main text.
Figure~\ref{composites3} exemplifies the extent of composite anomalous $\mathrm{O}_{3}$ concentrations as a function of pressure at 200 (top left), 300 (top right), and 400~hPa (bottom left) for AR $\Theta_{IVT}$ bins 2, 4, and 6.
These are the same as Figure~\ref{composites2} in the main text, but for the other three bins in AR direction of travel.
AR directions are reported in the titles of each figure. The bottom right panel in each of the 2x2 figures indicates the vertical cross section over the white dashed line in each of the other three figures, thus providing three dimensional information for the anomalous $\mathrm{O}_{3}$ concentrations and the average vertical transport of the Atmospheric Rivers' water vapor content.
Also indicated is the average potential vorticity contours (gray) and the 380~K isosurface (white).
The left-hand set of 2x2 figures are for ARs with peak IVT~$\lesssim$659~kg~m$^{-1}$~s$^{-1}$ (in the first quartile) and the right-hand column is for those with peak IVT in the fourth quartile. These figures are provided here in the appendix for completeness and to save space, and reflect parallel information as the other four $\Theta_{IVT}$ bins. The Composite maxima values are summarized in Figure~\ref{composite_results} in the main text.

\begin{figure*}
\centering

    \includegraphics[width=5.5in]{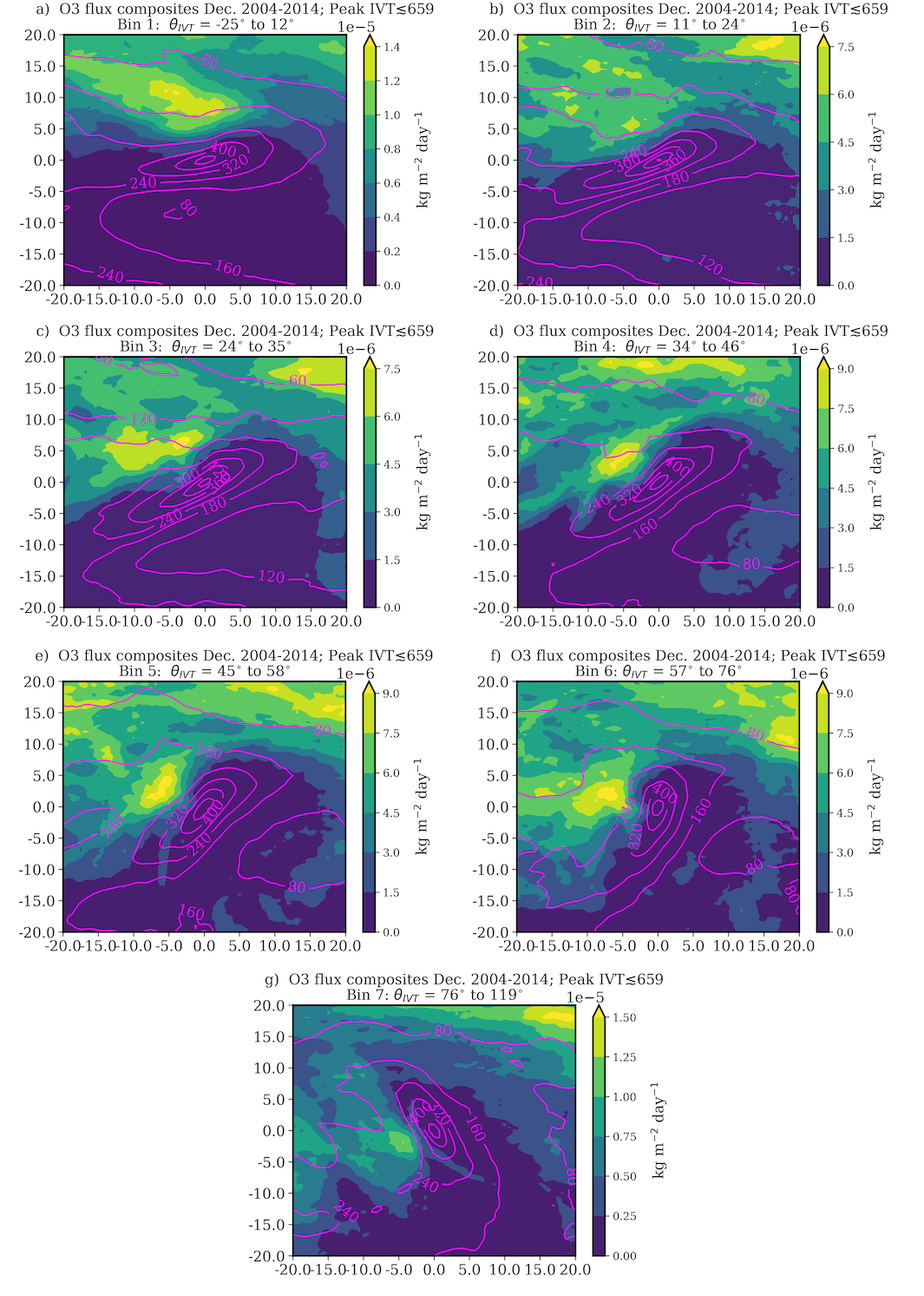}

\caption{ STT $\mathrm{O}_{3}$ flux for ARs with dry intrusion depths of P$\lesssim$350~hPa and AR peak IVT$\lesssim$659~kg~m$^{-1}$~s$^{-1}$. Figures generated with MERRA-2, and x and y axes are distance from the composite center in degrees longitude and degrees latitude, with x increasing in the east direction and y increasing in the poleward (north) direction.}
\label{fluxcomposites1}
\end{figure*}

\begin{figure*}
\centering

    \includegraphics[width=5.5in]{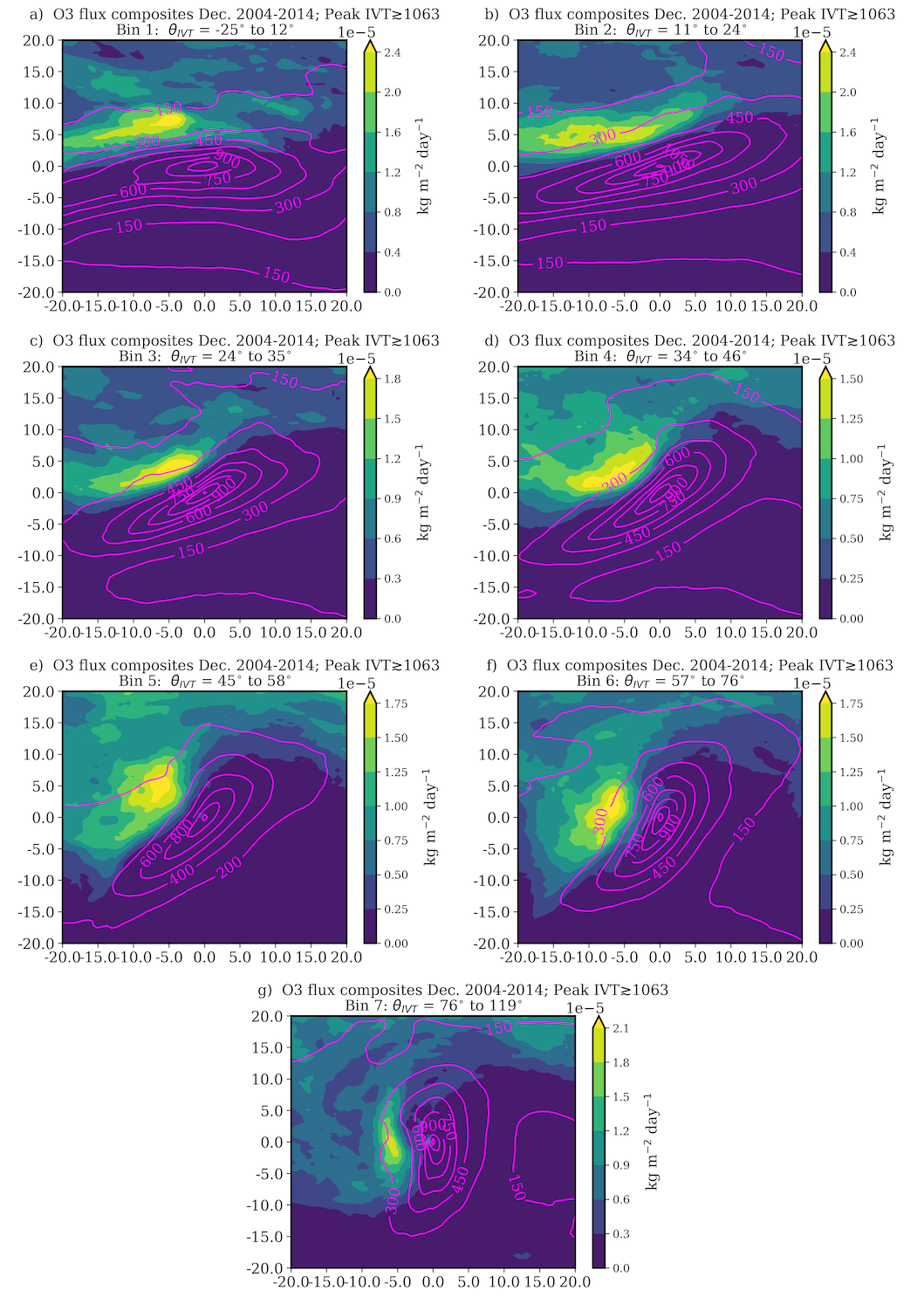}

\caption{ STT $\mathrm{O}_{3}$ flux for ARs with dry intrusion depths of P$\gtrsim$350~hPa and AR peak IVT$\gtrsim$1063~kg~m$^{-1}$~s$^{-1}$. Figures generated with MERRA-2, and x and y axes are distance from the composite center in degrees longitude and degrees latitude, with x increasing in the east direction and y increasing in the poleward (north) direction.}
\label{fluxcomposites2}
\end{figure*}

Figures~\ref{fluxcomposites1} and \ref{fluxcomposites2} show the $\mathrm{O}_{3}$ flux for ARs with dry intrusion depths with pressure greater than 350~hPa for ARs in bins of $\Theta_{IVT}$ and peak IVT (low and high for Figures~\ref{fluxcomposites1} and \ref{fluxcomposites2}, respectively). The figures demonstrate the smaller $\mathrm{O}_{3}$  flux for low IVT ARs, and larger flux for the more intense As.
The statistical summary for all events are plotted in Figure~\ref{o3flux}.

\section*{Open Research}

Data from the Modern-Era Retrospective analysis for Research and Applications, Version 2 instantaneous monthly \cite{Merra2_M} and 3-hourly \cite{Merra2_3hr} were used in the creation of this manuscript.
We also use $\mathrm{O}_{3}$ profiles from the Ozone Monitoring Instrument aboard the Aura spacecraft. The profiles and instructions for using them are available through the Aura validation center: \url{https://avdc.gsfc.nasa.gov/pub/data/satellite/Aura/OMI/V03/L2/OMPROFOZ/}.
All ARTMIP Tier 1 catalogs used in this work are available via the Climate Data Gateway at NCAR (DOI: \url{https://doi.org/10.5065/D62R3QFS}).
All data processing and plotting was done using Python version 3.8.12 installed with Miniconda3 \cite{conda}. For calculation of vertical cross sections, we use the publicly available package MetPy \cite{MetPy}.

\acknowledgments
The authors thank the reviewers of for their valuable feedback.
We are grateful to Emma Knowland for her insightful comments, which significantly improved the content and logical flow of this manuscript.
The authors were supported by the NASA OMI core science team Grant \#80NSSC21K0177 and the NASA Making Earth System Data Records for Use in Research Environments (MEaSUREs) Grant \#80NSSC18M0091.
K.R.H. acknowledges the support from the Schmidt Science Fellows, in partnership with the Rhodes Trust for the completion of this work.
Some computations for this paper were conducted at the Smithsonian High Performance Cluster (SI/HPC), Smithsonian Institution (https://doi.org/10.25572/SIHPC).
We also acknowledge the use of the ARTMIP project catalogs. ARTMIP is a grass-roots community effort and includes a collection of international researchers from universities, laboratories, and agencies.
Co-chairs and committee members include Jonathan Rutz, Christine Shields, L. Ruby Leung, F. Martin Ralph, and Michael Wehner, Ashley Payne, and Travis O'Brien. Details on catalogues developers can be found on the ARTMIP website.
ARTMIP has received support from the US Department of Energy Office of Science Biological and Environmental Research (BER) as part of the Regional and Global Climate Modeling program, and the Center for Western Weather and Water Extremes (CW3E) at Scripps Institute for Oceanography at the University of California, San Diego.


%
%

\bibliography{khall_atmosphere}

%
%
%
%
%

\end{document}